\RequirePackage{fix-cm}
\documentclass[smallextended,natbib,sort+compress]{svjour3}       
\smartqed  
\usepackage{graphicx}
\graphicspath{{./figures/}}
\usepackage{mathptmx}      
\usepackage{float}
\usepackage[dvipsnames]{xcolor}
\usepackage{xspace}
\usepackage{amsmath}
\usepackage{tikz}
\usetikzlibrary{positioning}
\usepackage{subcaption}
\usepackage{booktabs}
\usepackage[hidelinks]{hyperref}
\newcommand{\tbd}{\textcolor{red}{TBD}\xspace}
\newcommand{\supp}{\textcolor{black}{Supplementary Information}\xspace}
\newcommand{\uscsec}[2]{#1~U.S.C. §~#2}
\newcommand{\uscchap}[2]{#1~U.S.C. Ch.~#2}
\journalname{Submitted}
\begin{document}

\title{Law Smells
}
\subtitle{Defining and Detecting Problematic Patterns in Legal Drafting}


\author{%
	Corinna Coupette\textsuperscript{1,2} \and
	Dirk Hartung\textsuperscript{2,3} \and
	\mbox{Janis Beckedorf\textsuperscript{2,4} \and
	Maximilian B\"other\textsuperscript{5} \and
	Daniel Martin Katz\textsuperscript{2,3,6}}
}

\authorrunning{Coupette et al.} 

\institute{
	\textsuperscript{1}Max Planck Institute for Informatics, Saarbr\"ucken, Germany;
	\textsuperscript{2}Center for Legal Technology and Data Science, Bucerius Law School, Hamburg, Germany;
	\textsuperscript{3}CodeX -- The Stanford Center for Legal Informatics, Stanford Law School, CA, USA;
	\textsuperscript{4}Ruprecht-Karls-Universit\"at Heidelberg, Heidelberg, Germany;
	\textsuperscript{5}Hasso Plattner Institute, University of Potsdam, Potsdam, Germany;
	\textsuperscript{6}Illinois Tech – Chicago Kent College of Law, Chicago, IL, USA.
}

\date{}

\maketitle

\begin{abstract}
Building on the computer science concept of \emph{code smells}, 
we initiate the study of \emph{law smells}, i.e., patterns in legal texts that pose threats to the comprehensibility and maintainability of the law. 
With five intuitive law smells as running examples---namely,
duplicated phrase, long element, large reference tree, ambiguous syntax, and natural language obsession---, 
we develop a comprehensive law smell taxonomy. 
This taxonomy classifies law smells by when they can be detected, which aspects of law they relate to, and how they can be discovered. 
We introduce text-based and graph-based methods to identify instances of law smells, 
confirming their utility in practice using the United States Code as a test case. 
Our work demonstrates how ideas from software engineering can be leveraged to assess and improve the quality of \emph{legal} code, 
thus drawing attention to an understudied area in the intersection of law and computer science and highlighting the potential of computational legal drafting.
\end{abstract}

%


\section{Introduction}
\label{sec:intro}

In modern societies, law is one of the main tools to regulate human activities.
These activities are constantly changing, and law co-evolves with them.
In the past decades, human activities have become increasingly differentiated and intertwined, 
e.g., in developments described as \emph{globalization} or \emph{digitization}.
Consequently, legal rules, too, have grown more complex, 
and statutes and regulations have increased in volume, 
interconnectivity, and hierarchical structure \citep{katz2020,coupette2021}.

A similar trend can be observed in software engineering, albeit on a much shorter time scale. 
The global codebase has grown exponentially in recent years \citep{yu2014}, 
with GitHub as the largest source code host alone accounting for $250$~million repositories between $2008$ and $2020$.\!\footnote{%
	According to \url{https://octoverse.github.com}, $60$~million repositories were created in $2020$.} 
Over the course of its growth, this codebase has become increasingly interconnected when viewed through the lens of network analysis \citep{lima2014}.
Naturally, software engineers have turned to technology to keep track of this development and manage code interdependencies. 
While the challenges for law and software engineering and the constraints within which these challenges must be addressed are not identical, 
both domains share three important characteristics:
\emph{Materially}, their subject matters, legal rules on the one hand and code fragments on the other hand, 
contain commands intended to control (human or machine) behavior in order to achieve specific outcomes.
\emph{Procedurally}, output creation in both law and software engineering is distributed in time and space, 
and thus, both domains are subject to the challenges of dynamic multi-agent systems.
\emph{Methodologically}, lawyers and software engineers alike use abstractions to accommodate problems that are not fully known in the present.

The similarities between the law and software engineering domains suggest possibilities for knowledge transfer between them.
In this paper, we explore one such possibility for the software engineering subfield of \emph{refactoring}.
Introduced into academic discourse by \citet{opdyke1990} and popularized by \citet{becker1999} \citep[Second Edition:][]{fowler2018},
refactoring ``describes a disciplined technique for restructuring an existing body of code, altering its internal structure without changing its external behavior.''\footnote{Cf.~\url{https://refactoring.com/}.}
In software engineering, 
refactoring is indispensable for ensuring software quality and maintainability, 
and it is also subject to vivid academic discourse, 
inspiring detailed analyses of large code bases and even dedicated conferences.\!\footnote{Cf.~\url{http://www.msrconf.org/}.}
This has resulted in an actionable understanding of how various variables---%
e.g., programming language, team size, project size, or commit size \citep{ray2014}, 
repetition and duplicated code \citep{lopes2017}, 
or component size and open source model \citep{stamelos2002}---%
impact code quality.
	
In this paper, we demonstrate how concepts from refactoring can be used in law, focusing on the example of \emph{code smells}.
At a high level, a code smell is a characteristic of (a part of) the source code that may indicate a deeper problem in its design and implementation, 
highlighting a need for refactoring \citep{tufano2015}.
We port this idea to the legal domain, 
introducing the concept of \emph{law smells}.
Law smells constitute the first step towards both (semi-)automatically \emph{detecting} problematic parts of existing codifications and 
(semi-)automatically \emph{improving} codification quality.
They also pave the path towards reproducible, \emph{quantitative} quality measures for collections of legal rules, 
which allow us to assess the overall \emph{law climate}
and facilitate automatic code review and quality assurance.

The remainder of this paper is structured as follows.
In Section~\ref{sec:relatedwork}, we present related literature from law and computer science, along with interdisciplinary work in the intersection of both fields.
We develop the concept of \emph{law smells}, guided by five illustrative examples, in Section~\ref{sec:theory}, 
and describe methods to detect law smells in Section~\ref{sec:methods}.
In  Section~\ref{sec:practice}, we demonstrate the utility of our methods for detecting five example smells, 
deliberately showcasing a variety of ways to present law smell detection results to end users.
We discuss the limitations of our approach along with opportunities for future work in Section~\ref{sec:discussion}, 
and conclude in Section~\ref{sec:conclusion}.

\section{Related Work}
\label{sec:relatedwork}
Given its interdisciplinary nature, our work relates to prior research in law, in computer science, and in the intersection of both fields.
In the legal domain, the question of what constitutes high-quality law is the realm of legislative theory \citep{noll1973}, 
sometimes also referred to as \emph{legisprudence} \citep{wintgens1999}.
Within this wider field, our work is most closely related to research on legislative drafting, which, inter alia, develops suggestions to mitigate law smells.
Therefore, our work should be read in light of the debate on drafting manuals and other guidelines for the legislative process \citep[cf.][]{vanlochem2010,xanthaki2010}. 
While there exist handbooks that cover a wide range of drafting techniques and best practices \citep[cf.][]{xanthaki2014},
our work helps detect where these can be applied to improve existing legal texts.
Our improvements do not focus on criteria such as coherence, precision \citep[cf.][]{karpen2008}, or ``plain" language \citep[cf.][]{butt2013}, 
which are subject to an adjacent debate \citep[cf.][]{mousmouti2012,xanthaki2011}.
Moreover, there is extensive doctrinal debate on the use of digital technology  
in the legislative process \citep[cf.][]{dorsey2014,sartor2016},
for which law smells in legislative documents are an example.

In computer science, this paper is naturally inspired by the work of \citet{fowler2018} as laid out in Section~\ref{sec:intro}.
As the survey by \citet{sharma2017} demonstrates, 
\emph{code smells} or \emph{software smells} are subject to extensive research,
including in-depth studies on their limitations, interactions, and relations to other concepts such as software design ideas \citep[cf.][]{yamashita2013,speicher2020}.
Since the transfer of code smells to the legal domain is in an early stage, 
no meta-analysis currently exists for law smells, and we are only beginning to understand their prevalence and impact.

While the use of some concepts from mathematics or computer science, 
such as symbolic or deontic logic \citep[cf.][]{allen1957,allen1980}, 
or domain-specific (markup) languages \citep[cf.][]{allen1995},
in legal drafting has been debated for several decades \citep[cf.][]{,saxon1982,ziegler1989},
few publications in the intersection of law and computer science have used ideas from software engineering in the legal domain.
Among these, the work by \citet{li2015} is most similar to ours, 
as the authors adapt four code quality metrics to legal texts in order to quantitatively assess the quality of the United States Code. 
The metrics they select differ from ours, however, and the results are not embedded in an overarching conceptual framework.
Also related to our topic is the study by \citet{scott2012}, 
which introduces fifteen types of legal programming errors in telecommunication law and discusses ways to resolve them.
While starting from a similar point of departure, 
our work is distinct in that it provides operational tools to detect the deficiencies of and advocates for solutions independent from specific programming paradigms.
Finally, the tools presented in this paper could be used as a building block for legal drafting software, 
whose potential has been explored in prior work \citep[cf.][]{moens2006,hafner2007}.
As such, our ideas relate to the grander vision of a computerized legal information (retrieval) system \citep{bing1988,liebwald2015}, 
which has been discussed for several decades, 
yet will probably never be fully developed \citep{bing2010}.
Our concept of law smells is much more narrow but directly applicable to existing corpora of legal texts.
Hence, unlike a computerized legal information (retrieval) system, 
its implementation does not require large infrastructure investments.

\section{Theory: What Are Law Smells?}
\label{sec:theory}

In this section, we make the notion of a \emph{law smell} more precise
(\ref{subsec:theory:definition}), illustrate it via examples (\ref{subsec:theory:examples}),
and develop a taxonomy of law smells that prove particularly problematic in practice (\ref{subsec:theory:taxonomy}).
For conciseness, we refer to the government parties involved in creating legal rules as \emph{lawmakers} (e.g., parliaments or agencies),
to the parties affected by such legal rules as \emph{lawtakers} (e.g., law firms, companies, and citizens),
to people with a legal education as \emph{lawyers} (e.g., judges, attorneys, and legal scholars),
and to the labeled, often nested units of legal texts as \emph{elements of law} (e.g., Titles, Chapters, Sections, and Subsections of the United States Code).

\subsection{Definition}
\label{subsec:theory:definition}

A \emph{code smell}, as understood by \citet{fowler2018}, 
is ``a surface indication that usually corresponds to a deeper problem in the system.''\footnote{Cf. \url{https://martinfowler.com/bliki/CodeSmell.html}.}
Making only small changes to this statement, we arrive at our working definition of a \emph{law smell}:

\begin{definition}[Law Smell]\label{def:law-smell}
	A \emph{law smell} is a surface indication that usually corresponds to a deeper problem in a legal system.
\end{definition}

To clarify what this means, we highlight the most important building blocks of our definition.\!\footnote{%
	Some of the following clarifications are adapted to the legal domain from clarifications found for the code smell definition at \url{https://martinfowler.com/bliki/CodeSmell.html}.}
First, a law smell is a \emph{surface indication}:
It can be easily identified, perhaps algorithmically,
from the words and the structure of a legal text---%
even with very limited semantic knowledge.
Second, the problem indicated by a law smell concerns \emph{a legal system},
which---regardless of how exactly we define it%
\footnote{Defining what a legal system is, or when it exists, is a classic question in legal theory, see, e.g.,~\cite{hart1961,raz1980,luhmann2011}.}%
---%
goes beyond the body of rules in which the problem is identified
(for example, a smell discovered in the text of a regulation can indicate a problem in the statute-making process).
Third, the indicated problem is a \emph{deeper problem},
that is, an aspect of the design of a legal system that affects its usability for at least one of its stakeholders---%
e.g., by making it harder to maintain (for lawmakers) or to navigate (for lawtakers).
Finally, a law smell only \emph{usually} indicates a problem, i.e.,
upon closer inspection, it can turn out to be harmless or even necessary, 
given the legal rules that need to be expressed.
From a practical perspective, law smells can tell us where we should look to improve the legal system (as lawmakers)
or where we should pay extra caution (as lawtakers).

So far, we have defined law smells in terms of their \emph{form} and their \emph{consequences}.
To capture their \emph{content},
we propose to characterize each law smell systematically, 
using the five categories laid out in Table~\ref{tab:smell-criteria}.
We refer to the resulting overview as a \emph{law smell profile}.
For a given law smell, a law smell profile sketches
(1) what the law smell is and (2) why it is problematic.
It summarizes (3) how we might detect the smell and (4) how we might mitigate it,
and gives (5) at least one illustrative example.

\begin{table}[t]
	\caption{Categories used in a law smell profile.}\label{tab:smell-criteria}
	\centering
	\begin{tabular}{rll}\toprule
		\bfseries \#&\bfseries Category&\bfseries Question\\\cmidrule{1-3}
		1&Description & What is it?\\
		2&Problem & Why is it problematic?\\
		3&Detection & How do we detect it?\\
		4&Mitigation & If we decide that it needs mitigation, how can we mitigate it?\\
		5&Example & For illustration, how does it manifest in practice?\\
		\bottomrule
	\end{tabular}
\end{table}

\subsection{Examples}
\label{subsec:theory:examples}

We now introduce some particularly intuitive example law smells,
which will accompany us throughout the paper.
To maximize comprehensibility for readers with computer science or legal backgrounds,
we choose law smells that can be derived both via adaptation of the code smells known from the software engineering literature \citep[e.g.][]{fowler2018}
and by appealing to the intuition and experience of lawyers.
As the primary objective is to build intuition, our discussion is deliberately brief, 
and we detail the structured profiles of our example law smells in the~\supp.
We provide a broader view of the law smell landscape in Section~\ref{subsec:theory:taxonomy},
elaborate on the critical question of \emph{law smell detection} in Section~\ref{sec:methods},
and assess the practical prevalence of our example smells in Section~\ref{sec:practice}.

\subsubsection{Duplicated Phrase}
\label{sssec:theory:dp}

\emph{Duplicated phrase} is the legal equivalent of the software engineering code smell \emph{duplicated code}.
Lawyers smell it when they get the feeling that a text is verbose and repetitive.
More formally, a duplicated phrase is a phrase above a specified \emph{length} that has more than a specified number of \emph{occurrences} in a legal text.
Here, we construe ``phrase'' broadly and allow phrases to have parameters.
That is, a \emph{phrase} is a nonempty sequence of terms,
where a term is either a token (i.e., a sequence of non-whitespace characters roughly corresponding to a word)
or a placeholder for an argument that is itself a phrase (e.g., ``not later than \{period\} after \{date\}'').

Duplicated phrases are problematic because they increase the risk of inconsistencies introduced by incomplete edits,
and because they communicate legal content inefficiently.
We can detect duplicated phrases using scalable variants of $n$-gram search, possibly after some preprocessing,
where, for intuitive results,
the length and occurrence thresholds should be adaptive and inversely correlated---%
i.e., the longer a phrase, the fewer occurrences should be required for it to qualify as duplicated
(see Section~\ref{sssec:methods:dp} for details).
To eliminate duplicated phrases, we can introduce and rigorously reuse named variables and definitions.

\subsubsection{Long Element}
\label{sssec:theory:le}

\emph{Long element} is a legal adaptation of the software engineering code smell \emph{long function}.
Lawyers smell it when they get lost in the text they are reading or have trouble discerning those parts of the text that are relevant to them.
More formally, a long element is an element of law containing text that is long as assessed by some absolute or relative measure (e.g., its number of tokens or its quantile in a token length distribution).
Long elements are problematic because they may indicate a lack of structure or abstraction in the legal text,
making the law both harder to read and more complicated to maintain.
We can detect long elements using domain-specific adaptations of outlier detection methods (see Section~\ref{sssec:methods:le} for details).
To eliminate long elements, we can move some of their text to other (preexisting or newly created) elements or rewrite the text itself,
e.g., by applying mitigation strategies for other law smells.

\subsubsection{Ambiguous Syntax}
\label{sssec:theory:as}

\emph{Ambiguous syntax} is a legal adaptation of the software engineering code smell \emph{mysterious name}, with pinches of \emph{repeated switches} and \emph{speculative generality}.
Lawyers smell it, inter alia, when they litigate over the meaning of commas or argue about whether an \emph{or} is inclusive or exclusive.
More formally, ambiguous syntax is the use of logical operators (e.g., \emph{and}, \emph{or}, and \emph{no(t)}), control flow operators (e.g., \emph{if}, \emph{else}, and \emph{while}),
or punctuation (e.g., commas and semicolons) in a way that leaves room for interpretation.
Ambiguous syntax is problematic because it creates legal uncertainty,
which is often removed through costly lawsuits and sometimes leads lawmakers to adopt \emph{mathematically redundant syntax} like \emph{and/or}.\!\footnote{%
For an example of costly syntactic ambiguity that received plenty of media coverage in $2018$, see O'Connor v. Oakhurst Dairy -- 851 F.3d 69 (1st Cir. 2017), 
where the opinion of the First Circuit starts with ``For want of a comma, we have this case''.
In defense of the mathematically redundant \emph{and/or}, see \citet{robbins2017}.
}
We can detect instances of potentially ambiguous or mathematically redundant syntax using pattern matching with regular expressions (see Section~\ref{sssec:methods:as} for details).
To eliminate truly ambiguous instances,
logical operators can be used with their mathematical meaning,
\emph{xor} can be introduced as a shorthand for the exclusive \emph{or} (reserving \emph{or} for the cases with inclusive meaning),
and brackets can be added as syntax (e.g., to clarify operator binding).

\subsubsection{Large Reference Tree}
\label{sssec:theory:lrt}

\begin{figure}[t]
	\centering
	\includegraphics[width=0.9\linewidth]{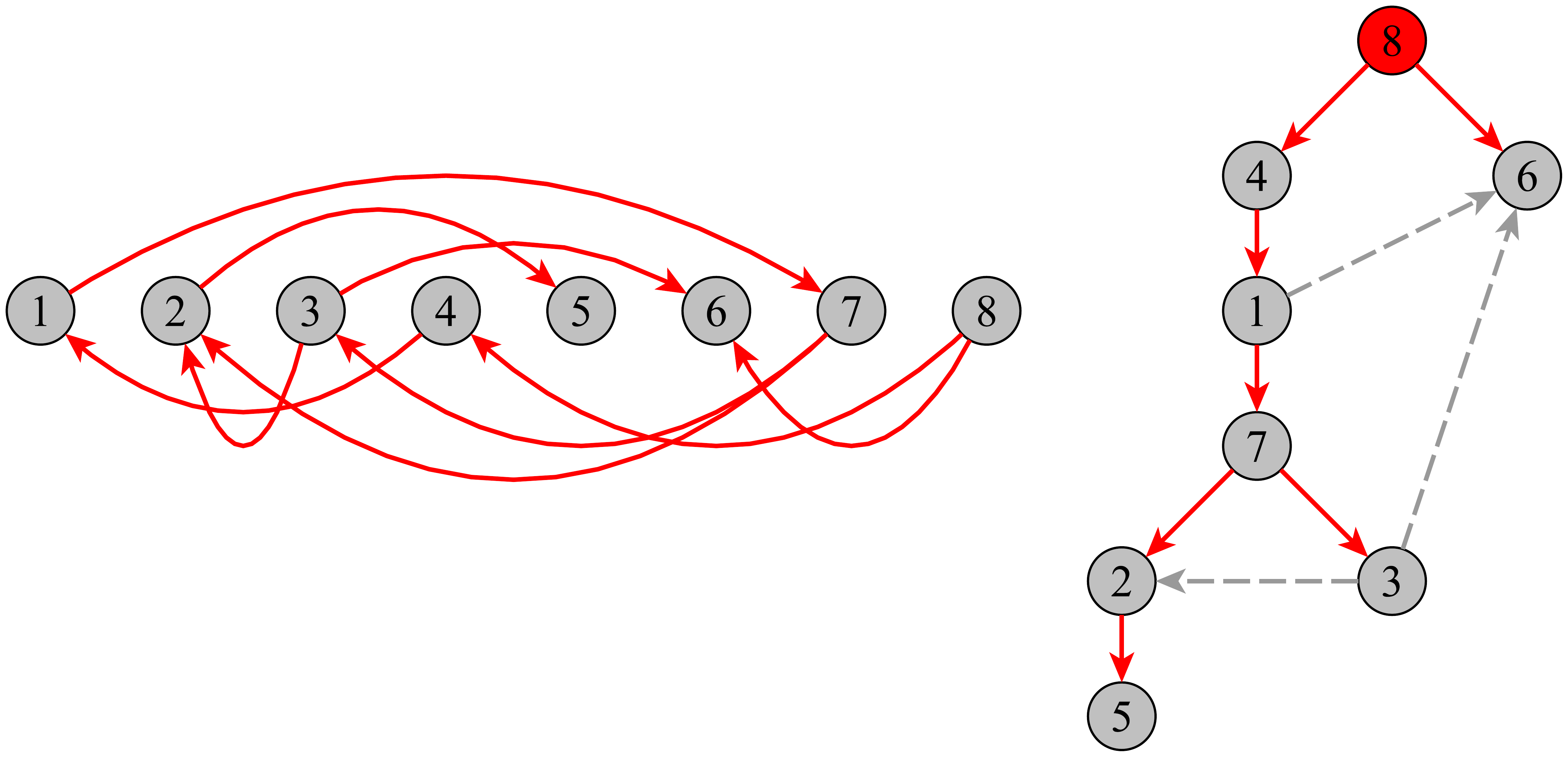}
	\caption{%
		Toy elements of law connected by reference edges (left) and a reference tree rooted at element $8$ (right), where edges not in $E_8$ are drawn as gray dashed arrows.
	}\label{fig:reference-tree}
\end{figure}

\emph{Large reference tree} is a legal adaptation of the software engineering code smell \emph{message chain}.
Lawyers smell it when they find themselves following many references to understand the full meaning of the text they were originally interested in.
More formally, a \emph{reference tree} rooted at an element of law $r$ is a tuple $T_r = (V_r, E_r)$,
where $V_r$ is the set of elements of law reachable from $r$ by following references (including $r$),
and $E_r$ is a minimal set of edges (references), which is generally not unique, 
such that each element of $V_r$ can be reached from $r$.
To illustrate this definition, a sample reference tree is depicted in Figure~\ref{fig:reference-tree}.
A \emph{large reference tree} is a reference tree whose edge set exceeds a given size $x$,
which represents the largest acceptable number of references that we must follow to fully understand the content of a root element.
The simplest case of a large reference tree is a \emph{long reference chain}.
Large reference trees are problematic because they increase the cognitive load involved in navigating legal texts,
and they raise the risk of unintended normative side effects when elements deep in the tree hierarchy are changed.
We can detect large reference trees using domain-specific adaptations of graph traversal algorithms (see Section~\ref{sssec:methods:lrt} for details).
To eliminate large reference trees, we can restructure the texts and references contained in its elements,
and we can make all references as specific as possible.

\subsubsection{Natural Language Obsession}
\label{sssec:theory:nlo}

\emph{Natural language obsession} is a legal adaptation of the software engineering code smell \emph{primitive obsession}.
Lawyers smell it, inter alia, when they are interested in the law related to a specific named entity (e.g., a committee of the United States Senate),
or when they try to automate their clients' compliance.
More formally, natural language obsession is the representation of typed data as natural language text, often without a standardized format.
It is problematic because typed data is notoriously hard to extract or maintain when represented using (inconsistently formatted) natural language.
We can detect natural language obsession using Named Entity Recognition methods (including methods as simple as pattern matching with regular expressions),
potentially augmented or validated by techniques detecting duplicated phrases (see Section~\ref{sssec:methods:nlo} for details).
To eliminate natural language obsession, we can create a data layer that is separate from the text (representation) layer,
use strong types for named entities,
and associate type checking, highlighting, and data analysis rules with the introduced types.

\subsection{Taxonomy}
\label{subsec:theory:taxonomy}

We now abstract and expand the ideas introduced above into a taxonomy of law smells.
To this end, we first derive the necessary distinctions
and then give more details on ten additional smells that we position within the resulting taxonomy.
This taxonomy is not meant to be exhaustive.
Rather, our goal is to provide a starting point for conceptual and empirical validation, discussion, and further refinement.

\subsubsection{Distinctions}
\label{subsubsec:taxonomy:distinctions}

To develop our taxonomy, we observe some commonalities and differences between the exemplary law smells introduced in Section~\ref{subsec:theory:examples}.

First, all of the law smells from Section~\ref{subsec:theory:examples} can be sniffed from the text only (in programming language parlance: they can be detected at \emph{compile time}; in legal theory parlance: they concern \emph{law in books}), 
but there might also be law smells that can only be detected when the law is used (i.e., at \emph{runtime}, concerning \emph{law in action}).
Thus, we distinguish \emph{static} and \emph{dynamic} law smells.

Second, we find that the \emph{duplicated phrase} and \emph{natural language obsession} smells are strongly related to the content of legal rules, i.e., the \emph{data} of the law,
while the \emph{ambiguous syntax} smell is related to how this content is expressed, i.e., the \emph{grammar} of the law,
and the \emph{long element} and \emph{large reference tree} smells relate to how the legal content is organized, i.e., the \emph{structure} of the law.
Consequently, we differentiate between \emph{data-related}, \emph{grammar-related}, and \emph{structure-related} law smells.
Note that since grammar-related law smells are generally static,
there is no need for this category in the dynamic setting.
Instead, we require an additional dynamic category, called \emph{relevance-related},
to gather smells related to elements of law with limited overall utility
(including, but not limited to, \emph{legal cleaning},
i.e., identifying and removing elements of law that are no longer relevant).

Third, while \emph{long element} (when measured in absolute terms, on which we focus here), 
\emph{ambiguous syntax}, and \emph{natural language obsession} can often be detected by looking at a single element of the law,
\emph{duplicated phrase} and \emph{large reference tree} (as well as \emph{long element} when measured in relative terms) 
mostly require looking at multiple elements of law together.
Therefore, we further distinguish \emph{local} (i.e., locally detectable) and \emph{non-local} (i.e., locally undetectable) law smells.

\begin{figure}[t]
	\centering
\begin{tikzpicture}[
	node distance=2cm and 0.5cm,
	ar/.style={-,>=latex},
	mynode/.style={
		text centered,
		text width=1.5cm,
		align=center
	},
	mybignode/.style={
		text centered,
		text width=5cm,
		align=left
	}
	]
	\node (root) {Law Smells};
	
	\node[mynode, above right=3cm and 0.5cm of root] (static) {static};
	
	\node[mynode, above right=of static] (stdata) {data-related};
	\node[mybignode, above right=0.25cm and 0.5cm of stdata] (dp) {\bfseries\itshape Duplicated Phrase};
	\node[mybignode, below right=0.25cm and 0.5cm  of stdata] (nlo) {\bfseries Natural Language Obsession};
	
	\node[mynode, right =of static] (ststructure) {structure-related};
	\node[mybignode, above right=0.25cm and 0.5cm  of ststructure] (le) {\bfseries Long Element};
	\node[mybignode, above right=-0.15cm and 0.5cm of ststructure] (lrl) {Long Requirements List};
	\node[mybignode, right=of ststructure] (rc) {\bfseries\itshape Large Reference Tree};
	\node[mybignode, below right=-0.15cm and 0.5cm of ststructure] (fe) {\itshape Rule Envy};
	\node[mybignode, below right=0.25cm and 0.5cm of ststructure] (fi) {\itshape Failed Integration};
	
	\node[mynode, below right =of static] (stgrammar) {\textsc{grammar-related}};
	\node[mybignode, above right=0.25cm and 0.5cm of stgrammar] (as) {\bfseries Ambiguous Syntax};
	\node[mybignode, right=of stgrammar] (cl) {Complicated Logic};
	\node[mybignode, below right=0.25cm and 0.5cm of stgrammar] (ie) {\itshape Inconsistent Enumeration};
	
	\node[mynode, below right=3cm and 0.5cm of root] (dynamic) {dynamic};
	
	\node[mynode, above right=of dynamic] (dydata) {data-related};
	\node[mybignode, right=of dydata] (to) {\itshape Term Overloading};
	
	\node[mynode, right =of dynamic] (dystructure) {structure-related};
	\node[mybignode, above right=0.25cm and 0.5cm of dystructure] (dc) {Divergent Change};
	\node[mybignode, right=of dystructure] (ss) {\itshape Shotgun Surgery};
	\node[mybignode, below right=0.25cm and 0.5cm of dystructure] (or) {\itshape Overbroad Reference};
	
	\node[mynode, below right=of dynamic] (dygarbage) {\textsc{relevance-related}};
	\node[mybignode, right=of dygarbage] (lae) {\itshape Lazy Element};
	
	\draw 
	(root) -- (static);
	\draw 
	(stdata) -- (static);
	\draw 
	(stdata) -- (dp.west);
	\draw 
	(stdata) -- (nlo.west);
	\draw 
	(ststructure) -- (static);
	\draw 
	(ststructure) -- (le.west);
	\draw 
	(ststructure) -- (lrl.west);
	\draw 
	(ststructure) -- (rc.west);
	\draw 
	(ststructure) -- (fe.west);
	\draw 
	(ststructure) -- (fi.west);
	\draw 
	(stgrammar) -- (static);
	\draw 
	(stgrammar) -- (as.west);
	\draw 
	(stgrammar) -- (cl.west);
	\draw 
	(stgrammar) -- (ie.west);
	\draw 
	(root) -- (dynamic);
	\draw
	(dydata) -- (dynamic);
	\draw 
	(dydata) -- (to.west);
	\draw 
	(dystructure) -- (dynamic);
	\draw 
	(dystructure) -- (dc.west);
	\draw 
	(dystructure) -- (ss.west);
	\draw 
	(dystructure) -- (or.west);
	\draw
	(dygarbage) -- (dynamic);
	\draw 
	(dygarbage) -- (lae.west);
\end{tikzpicture}
	\caption{Depiction of the law smell taxonomy derived in the main text.
		Smells from Subsection~\ref{subsec:theory:examples} are printed in bold,
		and primarily non-local smells are typeset in italics.
		The names of subcategories that differ across static and dynamic smells are emphasized using small caps.
	}\label{fig:law-smell-taxonomy}
\end{figure}
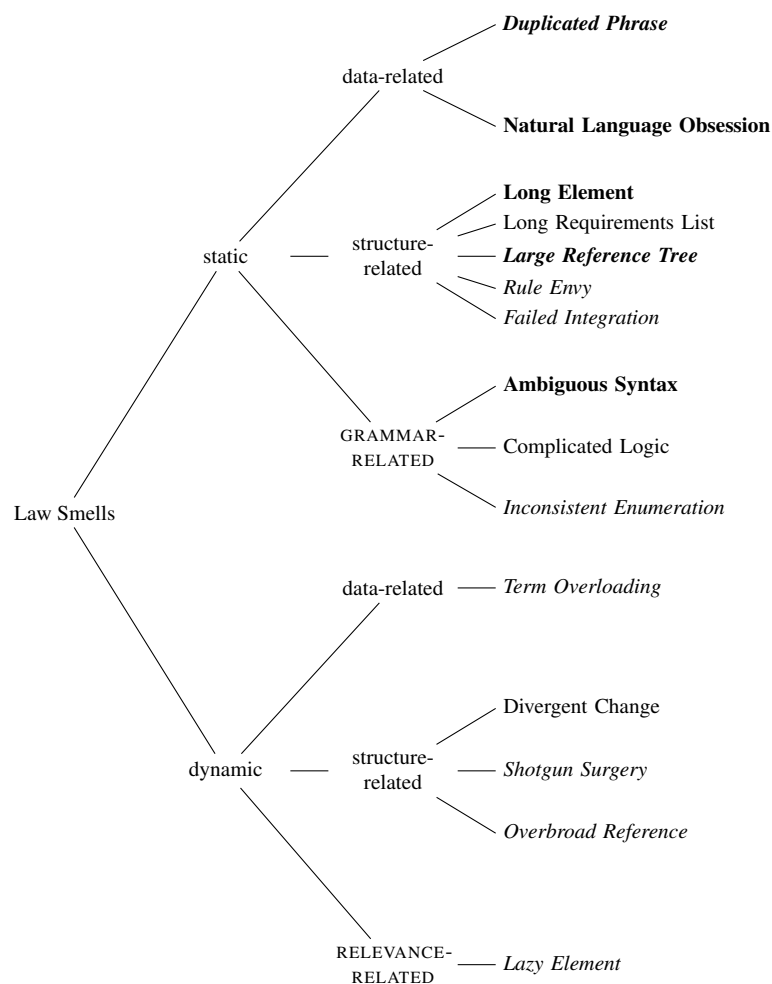

Table~\ref{tab:taxonomy-splits} summarizes the distinctions just introduced,
and Figure~\ref{fig:law-smell-taxonomy} gives an overview of the resulting taxonomy.
Figure~\ref{fig:law-smell-taxonomy} also contains ten new law smells in addition to those already classified above, 
thus painting a more complete picture of the law smell landscape we envisage.

\begin{table}[t]
	\centering
	\caption{Overview of distinctions used in the law smell taxonomy.}\label{tab:taxonomy-splits}
	\begin{tabular}{rp{0.2\linewidth}p{0.5525\linewidth}}\toprule
		\bfseries Distinction&\bfseries Categories&\bfseries Question\\\cmidrule{1-3}
		 Stability &static, dynamic& Can it be detected by looking at the text alone?\\
		 Modality &data, structure,\newline grammar, relevance& What aspect of the law does it relate to?\\
		 Locality &local, non-local& Can it be detected by looking at a single element of law?\\\bottomrule
	\end{tabular}
\end{table}

\subsubsection{Details}
\label{subsubsec:taxonomy:details}

We now elaborate on the new law smells from Figure~\ref{fig:law-smell-taxonomy}.
Of these smells, five are static and five are dynamic,
and we discuss them in the order in which they occur in Figure~\ref{fig:law-smell-taxonomy}.

\emph{Long requirements list},
a legal adaptation of the software engineering code smell \emph{long parameter list},
is a static law smell describing the situation in which a stated legal consequence only applies given a large number of preconditions.
As preconditions might have their own preconditions to be resolved recursively,
computer scientists may think of the setting as a search tree whose initial branches need to be expanded until we have either a negative answer on one branch or positive answers on all branches.
The efficient lawyer's task is to pick,
using their legal intuition and the facts of an individual case,
an expansion order that minimizes the number of search nodes to be expanded.
This can only work if they can retain at least the initial preconditions in short-term memory, and therefore,
the cognitive load imposed by these preconditions should not exceed the average lawyer's---if not the average lawtaker's---%
short-term retention capacity.
Although the precise measurement of short-term memory is still debated in cognitive psychology, for the purposes of flagging long requirements lists,
we can still use a conservative maximum number of requirements $x$ as a heuristic.
To shorten long requirements lists, lawmakers can introduce new concepts bundling requirements that cooccur frequently.

\emph{Rule envy}, a legal adaptation of the software engineering code smell \emph{feature envy},
describes a setting in which a legal rule is more closely related to rules from other parts of the law than to the rules with which it is grouped (e.g., the rules in its own section).
Typical warning signs that an element of law is affected by rule envy are its extensive (implicit or explicit) references to rules from far-away parts of the law,
paired with an absence of (implicit or explicit) references to rules in its immediate neighborhood.
Rule envy is problematic because it goes against the rationale that underlies the structuring of law into separate elements:
improving its comprehensibility and maintainability by grouping related content together.
\emph{In theory}, rule envy can be mitigated by placing the envious element with the elements it envies.
However, \emph{in practice}, many elements of law rely on legal rules from different places, and consequently, it may not always be clear where an element naturally belongs.
Furthermore, some prominent drafting techniques such as \emph{index sections} deliberately create rule envy,
and when used with caution, the gain in navigational efficiency associated with these techniques may outweigh the drawbacks of the resulting law smell.

\emph{Failed integration} can be conceived as the group-level opposite of \emph{rule envy}:
it refers to a sequence of (mostly lower-level) elements of law that are isolated from the environment in which they are placed---%
although their position within the structure of the law suggests otherwise.
As the train of thought spun in the environment of an affected sequence is interrupted by failed integration, it makes navigating the law unnecessarily difficult.
This smell frequently results, e.g., from hasty or unenthusiastic implementation of supranational legal rules into national legal systems 
(e.g., when the German legislator needs to implement law mandated by the European Union),
especially when supranational and national law differ in their legal concepts.
As a result, affected sequences often differ from their environment in drafting style and desired interpretation of core terms,
which increases, inter alia, the risk of \emph{inconsistent enumeration} and  \emph{term overloading} (described below).
Failed integration can be mitigated by either moving the affected sequence completely out of its environment,
i.e., making it into an independent unit of law,
or rephrasing its elements for better coherence with its environment.

\emph{Complicated logic} arises when legal rules contain nested conditionals, switches, or loops, often with elaborate heads including negations.
It sometimes cooccurs with \emph{long requirements list} and frequently results in \emph{ambiguous syntax}.
In legal texts just as in code, humans have trouble parsing complicated logic because it puts significant strain on their mental cache,
impairing the usability of the law for both lawmakers and lawtakers.
Lawmakers can follow programmers' mitigation principles to eliminate complicated logic,
e.g., they can decompose or consolidate conditionals where possible.
In extreme cases, e.g., when lawmakers communicate a table in writing (example: \uscsec{7}{8732}, which details loan rates per unit of crop per crop type and year),
the best choice is to use an entirely different structure to present the content (in the example: a table).

\emph{Inconsistent enumeration} refers to the unstandardized usage of enumeration labels (e.g., large Latin alphabets, small Latin alphabets, Arabic numbers, large Roman numerals, small Roman numerals, etc.) across nesting levels in legal texts (or, more mathematically: the lack of a bijection between enumeration labels and nesting levels).
As programming languages use the same syntax at all levels to indicate nesting (e.g., lists are nested simply by using list syntax inside list syntax) or enumeration (e.g., switch-case expressions in some languages),
this law smell has no close counterpart in the software engineering code smell literature.
Its repercussions, however, could be compared to those of overly lenient XML schema definitions:
The affected text becomes hard to parse, be it for humans or machines, because the meaning of an encountered nesting element can only be resolved within a larger context.
Lawmakers can mitigate inconsistent enumeration by (legally) mandating a one-to-one mapping between enumeration labels and nesting levels, 
which is then used consistently across legal documents and can, consequently, be relied upon by all stakeholders.

\emph{Term overloading}, a legal mixture of the software engineering code smells \emph{mysterious name} and \emph{mutable data},
occurs when the same term is used to denote different concepts in different places,
either explicitly (\emph{explicit namespacing}, e.g., using \emph{scoping language} such as ``For the purposes of this section, $\dots$'')
or implicitly (\emph{implicit namespacing}, e.g., through jurisprudence).
It is classified as a dynamic smell because divergent meanings of the same term often result from the interpretation of legal rules in practice, e.g., by courts.
Term overloading increases the cost of navigating the law for lawtakers,
as the meaning of a term becomes context-dependent,
and the context-applicable meaning becomes accessible only through (often non-local) further reading (explicit namespacing) or background knowledge (implicit namespacing).
Additionally, it raises maintenance costs for lawmakers, especially when tasked with applying changes to all instances of a concept labeled by an overloaded term.
Term overloading can be mitigated by introducing separate labels for each concept referenced by the overloaded term,
and by using strategies similar to those combating \emph{natural language obsession}.

\emph{Divergent change}, a legal adaptation of the eponymous software engineering code smell,
describes a situation in which the same element of law needs to be changed in different ways for different reasons.
For example, an element holding a sequence of generic legal rules is affected by divergent change if in different application contexts,
different specific exceptions need to be made to rules contained in different sequence parts,
and the lawmaker's original strategy is to place the exceptions with the generic provisions.
Divergent change is a dynamic smell because it can be detected only during maintenance.
It indicates that the chosen level of abstraction is inadequate for the affected legal rules,
thus pointing to a problem in the structure of the law that may also impair its usability (e.g., the affected element of law may lack conceptual clarity or be hard to navigate).
Divergent change can be counteracted by restructuring the code to enforce a clear separation of concerns between application contexts,
such that the rules that are truly shared between contexts are separated from the context-specific details.

\emph{Shotgun surgery}, like the eponymous software engineering code smell, is the opposite of (the legal adaptation of) divergent change.
It captures the setting in which one high-level change to the law necessitates many small changes to legal rules in many different parts of the law,
and as such, it often indicates \emph{natural language obsession}.
Just as divergent change,
shotgun surgery is a dynamic smell because it can only be detected during maintenance.
Shotgun surgery is not only slow but also hard for lawmakers to perform consistently (and painful for lawtakers to follow).
As a result, conducting shotgun surgery may lead to inconsistencies in the law.
This makes future amendments to the law even harder, if not impossible, to perform correctly,
and produces unwarranted legal uncertainty (and litigation costs associated with removing that uncertainty) for lawtakers confronted with the resulting inconsistencies.
Where it is due to natural language obsession, shotgun surgery can be counteracted by the strategies used to neutralize that smell.
Otherwise, close inspection of those parts of the law that are altered by shotgun surgery might yield characteristics that could be abstracted into new concepts or generalized rules for smell mitigation.

An \emph{overbroad reference} is a reference to a structural element of law that is more general than its use case requires.
It is a legal adaptation of the software engineering code smell \emph{speculative generality},
which we classify as dynamic because the use case requirements of a legal rule often become clear only as it is used in practice.
This smell typically manifests in formulations such as ``Subject to the provisions of X [related to Y]'', where X is a large structural element of the law,
e.g., a Title, a Part, or a Chapter of the United States Code,
and Y (if present) is some roughly sketched topic.
Overbroad references shift the cost of identifying the \emph{relevant} referenced material from lawmakers to lawtakers,
decreasing efficiency and legal certainty and increasing the probability of errors.
Lawmakers can eliminate overbroad references by identifying the applicable legal rules themselves and referencing these rules as precisely as possible.
Their need for using overbroad references to hedge against future changes could be diminished by using computational tools to track the dependencies among legal rules.

\emph{Lazy element}, a legal adaptation of the eponymous software engineering code smell,
refers to elements of law that are hardly relevant in practice.
It is a dynamic smell because both laziness and non-laziness can develop over time:
For example, a legal rule can be relied upon by a lot of other rules when it is first promulgated,
but newer rules can gradually supersede it,
or it can be completely isolated at the start but then form the basis of many lawsuits or contracts.
The most extreme case of lazy elements is \emph{dead law}:
legal rules that are simply not needed (anymore)---%
e.g., because their preconditions imply that a person needs to have witnessed a certain event that happened $x$ years ago,
and $x$ exceeds maximum observed human lifespan.
In the absence of systematic compression or cleaning strategies for law (sunset clauses are still a fringe phenomenon),
lazy elements cause bloating,
and bloated law is frustrating to read and maintain.
Lawmakers can eliminate lazy elements by integrating them with their more industrious peers or, if the elements turn out to be dead, removing them entirely (through \emph{cleanup acts} such as the \emph{Rechtsbereinigungsgesetze} in Germany).

\section{Methods: How Do We Detect Law Smells?}
\label{sec:methods}

In this section, we introduce methods to detect the law smells presented in Section~\ref{sec:theory}.
After clarifying our assumptions regarding the input we are working with (\ref{subsec:methods:assumptions}),
we assemble our law smell detection toolkit (\ref{subsec:methods:tools})
and then provide more detail on the tools to detect our example law smells (\ref{subsec:methods:examples}).
Many of the techniques we build on are simple by themselves,
but intricacies arise when adapting them to tackle the \emph{law smell detection problem}.
For the sake of clarity, we define this problem as follows.

\begin{definition}[Law Smell Detection Problem]\label{def:law-smell-detection}
	Given a set of legal documents with associated metadata,
	 identify instances of law smells in elements of the set.
\end{definition}

Here, \emph{set} is used in the mathematical sense,
and the terms \emph{legal document} and \emph{associated metadata} are deliberately open.
For example, the legal documents could be different versions of all Titles of the United States Code,
and metadata could specify versioning relationships between them.
Note that we do not require the documents to be \emph{amendable},
but some dynamic law smells can only occur in amendable documents.

\subsection{Assumptions}
\label{subsec:methods:assumptions}

Which methods are available for detecting law smells
depends heavily on how the law (used in the following to denote the legal document set of interest) is stored and accessed.
If the law is accessed in printed form,
manual detection is the only method that is immediately applicable,
and there are generally no guarantees that manual detection is performed consistently or documented comprehensively.
If the law is stored in a version-controlled database with rich semantic annotations and a plethora of systematic indices,
automatic detection of law smells can be performed consistently and documented comprehensively by the database administrator
using mostly adaptations of standard software engineering and database management tools.
But while the latter state certainly seems desirable,
most use cases for law smell detection lie between the sketched extremes:
Lawmakers and lawtakers access the law in a semi-structured format, e.g., HTML or XML,
with hardly any access to systematic indices or semantic annotations.
Therefore, in the following, we focus on methods to detect law smells automatically or semi-automatically,
given access to the law in a semi-structured format.
We also assume that all legal documents provided in this format contain \emph{text},
which can be segmented into \emph{tokens}.

\begin{figure}[t]
	\centering
	\includegraphics[width=0.9\linewidth]{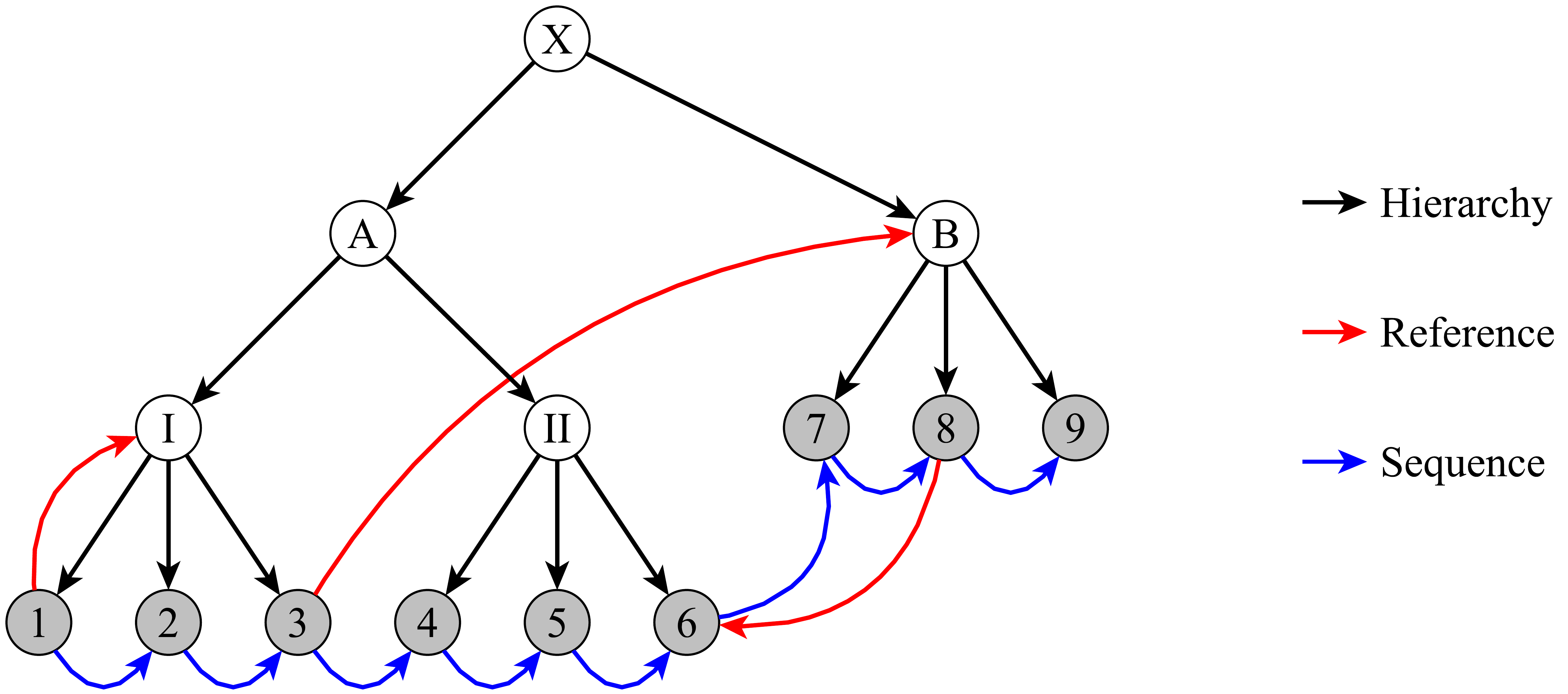}
	\caption{Structure types typically found in legal documents, illustrated using a toy document X.
		X consists of nine sequence-level elements (gray)
    	containing text (e.g., sections),
		which are connected via sequence edges (blue).
		These are nested inside two higher-level structures (A and B, e.g., parts),
		the first of which holds two additional mid-level structures (I and II, e.g., chapters),
		and all elements are connected by a tree of hierarchy edges (black).
		Finally, the text in some elements references other elements (not necessarily on the same level),
		and these references are captured by reference edges (red).
		Note that the sequence-level elements can have further substructures (e.g., subsections),
		which can also reference or be referenced by other elements.
	}\label{fig:law-structure-types}
\end{figure}

We make additional assumptions regarding the information that can be extracted from the semi-structured format (e.g., via preprocessing).
These assumptions strike a balance between generality (to ensure that they hold in a large number of jurisdictions and for a large number of document types)
and specificity (to ensure that they provide useful constraints for solving the law smell detection problem).
They are encapsulated in the legal document model developed by \cite{coupette2021}, which builds on ideas introduced by \cite{katz2020}.
In brief, this model assumes that legal documents,
in addition to their text,
contain at least three types of \emph{structure}:
\emph{hierarchy}
(elements of law are nested),
\emph{sequence} (most text is placed in elements that are uniquely labeled and sequentially ordered, and the labels of \emph{sequence-level elements} are used as the primary index for text retrieval),
and \emph{reference} (text in one element can contain cross-references or citations to other elements).
For example, the United States Code is structured into \emph{Titles} as the top level of a \emph{hierarchy} containing substructures such as \emph{Parts}, \emph{Subparts}, \emph{Chapters}, \emph{Subchapters}, and \emph{Sections},
its \emph{sequence} is indicated by section labels,
and \emph{reference} occurs to elements on all levels of the hierarchy.
Similar observations apply to many collections of rules,
be it from other branches of government (e.g., the Code of Federal Regulations),
in other countries (e.g., Germany),
or on other regulatory levels (e.g., European Union law).
Note that all assumed structure types can be captured by graphs
(ideally, directed multigraphs with one edge type per structure type),
and that the tree induced by the hierarchy edges need not be regular or balanced,
as illustrated in Figure~\ref{fig:law-structure-types}.
This facilitates the development of our law smell detection toolkit, which we describe next.

\subsection{Tools}
\label{subsec:methods:tools}

Since legal documents contain \emph{text} with structure that can be represented as a \emph{graph},
our law smell detection tools naturally fall into two categories:
\emph{text-based} and \emph{graph-based}.
Text-based tools comprise law-specific adaptations of methods from natural language processing and sequence mining (where the text is understood as a sequence of tokens or characters),
while graph-based tools encompass law-specific adaptations of graph algorithms and methods from legal network analysis.
Table~\ref{tab:smell-tools} groups the law smells from our taxonomy (cf. Figure~\ref{fig:law-smell-taxonomy}) according to these categories,
placing each smell where the emphasis of our detection strategies lies.
For each law smell in the \emph{text} category, we list the abstract task(s) associated with its detection,
and for each law smell in the \emph{graph} category, we list the detection-relevant edge types (hierarchy, sequence, or reference).

\begin{table}[t]
	\caption{Classification of law smells by primary detection method.
		Dynamic smells are typeset in italics;
		their detection may involve
		consulting external data.}\label{tab:smell-tools}
	\centering
	\begin{subtable}{\linewidth}
		\subcaption{Text-based detection}
		\centering
		\begin{tabular}{p{0.353\linewidth}p{0.575\linewidth}}\toprule
			\bfseries Law Smell&\bfseries Relevant Abstract Task(s)\\\cmidrule{1-2}
			Duplicated Phrase &Pattern Detection (Unknown Patterns)\\
			Natural Language Obsession &Pattern Detection (Known/Unknown Patterns)\\
			Long Element&Segmentation, Counting\\
			Long Requirements List &Pattern Detection (Known Patterns), Counting\\
			Ambiguous Syntax &Pattern Detection (Known Patterns)\\
			Complicated Logic &Pattern Detection (Known Patterns)\\
			\emph{Term Overloading} &Pattern Detection (Known/Unknown Patterns)\\
			\emph{Divergent Change} &Diffing\\
			\emph{Shotgun Surgery} &Diffing\\\bottomrule
			
		\end{tabular}
	\end{subtable}\\[12pt]
	\begin{subtable}{\linewidth}
		\subcaption{Graph-based detection}
		\centering
		\begin{tabular}{p{0.353\linewidth}p{0.575\linewidth}}\toprule
			\bfseries Law Smell&\bfseries Relevant Edge Type(s)\\\cmidrule{1-2}
			Large Reference Tree &Hierarchy, Reference\\
			Rule Envy &Sequence, Reference\\
			Failed Integration &Hierarchy, Sequence, Reference\\
			Inconsistent Enumeration &Hierarchy\\
			\emph{Overbroad Reference} &Hierarchy, Reference\\
			\emph{Lazy Element} &Reference\\\bottomrule
		\end{tabular}
	\end{subtable}
\end{table}

Among the tasks relevant for text-based detection,
\emph{segmentation}, \emph{counting}, and \emph{diffing} are relatively simple \emph{conceptually}, but the devil lies in the details.
Here, the primary challenges are choosing adequate inputs and making sensible judgment calls when translating raw outputs into law smell detection results.
For the \emph{detection of known patterns}, our Swiss army knives are Regular Expressions (Regex),
as they allow us to incorporate legal background knowledge easily
and their results are completely interpretable.
For some of the grammar-related smells (e.g., \emph{ambiguous syntax} or \emph{long requirements list}),
detection could further rely on parse trees or other text representations containing syntactic information,
provided they are deemed sufficiently reliable.
Additionally, Named Entity Recognition (NER) tools could be used whenever the known patterns of interest are named entities according to their definition.

For the \emph{detection of unknown patterns},
what methods we can choose from depends on how we define a \emph{pattern}:
Is a pattern a sequence of tokens that is repeated in some sense \emph{frequently}---%
or is it something different (e.g., a characteristic sequence of part-of-speech tags)?
If the former, is frequency assessed using some fixed (absolute or relative) threshold ($\rightarrow$~frequent pattern mining),
is it determined relative to our expectations under a certain null model  ($\rightarrow$~statistical pattern mining),
or is it based on whether replacing the pattern with a new symbol aids compression ($\rightarrow$~information-theoretic approaches)?
Should we allow a pattern to contain gaps or mismatches ($\rightarrow$~no traditional $n$-grams),
to contain or overlap with other patterns ($\rightarrow$~no greedy assignment approaches),
or even to have parameters (e.g., P(term, reference) $=$ ``\{term\} has the meaning given such term in \{reference\}'')?
Our answers to these questions should be guided by the law smell of interest (e.g., we might require exact matches for \emph{duplicated phrase})
and the characteristics of the documents in which we would like to detect it
(e.g., the importance of algorithm scalability increases with the length of the document at hand).

When performing graph-based law smell detection,
the biggest challenge lies in finding an appropriate resolution level at which to represent the edges of interest.
This is a common problem in legal network analysis \citep[cf.][]{coupette2021,katz2020},
and its solutions are highly context-dependent.
Once the edge resolution level has been determined,
most graph-based law smell detection methods can be built from relatively simple primitives,
such as walks along edges and set operations on node neighborhoods.
In specific cases, prepackaged graph algorithms might also be available
(e.g., given an intelligently constructed input graph, a graph clustering algorithm could be used to identify instances of \emph{failed integration}).
Here, we must take extra care to ensure that all assumptions made by the prepackaged algorithm (including those that are only \emph{implicit}) still hold in the scenario of interest.

\subsection{Examples}
\label{subsec:methods:examples}

To show the previous deliberations in action,
we now elaborate on our methods to detect the example law smells from Section~\ref{subsec:theory:examples}.
For concreteness, we focus on sets of legal documents that fulfill the assumptions specified in Section~\ref{subsec:methods:assumptions}
and contain \emph{codified statutes} or \emph{codified regulations}
(e.g., the United States Code, the Code of Federal Regulations,
or consolidated versions of German \emph{root laws} [\emph{Stamm\-ge\-setze}] and \emph{root regulations}  [\emph{Stammverordnungen}]).
To demonstrate that our methods work well in practice,
we use them to detect law smells in the United States Code in Section~\ref{sec:practice}.

\subsubsection{Duplicated Phrase}
\label{sssec:methods:dp}

Recall from Section~\ref{sssec:theory:dp} that a \emph{duplicated phrase} is a phrase above a specified length (\emph{maximum phrase length})
that has more than a specified number of occurrences in a legal text (\emph{maximum occurrence frequency}),
where a \emph{phrase} is a nonempty sequence of terms,
each of which is either a token or a placeholder for an argument that is itself a phrase.
We classify this smell as static, data-related, and non-local (cf.~Figure~\ref{fig:law-smell-taxonomy}),
with text-based detectability based on discovering unknown patterns  (cf.~Section~\ref{subsec:methods:tools}).

With this definition, the number of duplicated phrases in a legal document depends heavily
on how we treat the parameters \emph{maximum phrase length} and \emph{maximum occurrence frequency}.
Furthermore, if we choose the parameters to be constant,
duplicated phrases become practically \emph{downward closed}, i.e.,
any minimum-length subsequence of a duplicated phrase will also be a duplicated phrase.
For na\"ive parametrizations, we thus face problems similar to those encountered in frequent pattern mining (see~\citet{aggarwal2014} for an overview):
(1) If we choose the thresholds too high, we will miss duplicated phrases that we would like to include,
and if we choose them too low, we will catch phrases that we would like to exclude,
(2) there will be lots of redundancy in the result set, and
(3) reporting all our results will overwhelm the users of our method.

Thus, drawing inspiration from pattern set mining \citep[see, e.g.,][]{vreeken2011},
rather than identifying \emph{all} duplicated phrases,
our goal becomes to identify a \emph{set} of duplicated phrases whose refactoring we expect to yield the biggest text quality improvements.
To reach this goal, we use the \textsc{Dupex} algorithm introduced by \cite{coupette2021a}.
\textsc{Dupex} leverages information theory to identify duplicated phrases in legal documents, 
selecting sequences of tokens as duplicates based on the compression we achieve when replacing them (which results in adaptive length and occurrence frequency thresholds).
As input to \textsc{Dupex}, we feed a preprocessed version of the text we are interested in,
with common named entity patterns replaced by correspondingly labeled placeholders.
Hence, some of the duplicated phrases we discover are naturally parametrized.

\subsubsection{Long Element}
\label{sssec:methods:le}

Recall from Section~\ref{sssec:theory:le} that a \emph{long element} is an element containing legal text that is long as assessed by some absolute or relative measure.
We classify this smell as static, structure-related, and---emphasizing  absolute measurement---local (cf.~Figure~\ref{fig:law-smell-taxonomy}),
with text-based detectability based on solving segmentation and counting tasks (cf.~Section~\ref{subsec:methods:tools}).

When detecting long elements, we need to answer three questions:
\begin{enumerate}
	\item Which \emph{type of element} are we interested in
	(e.g., do we want to identify long Titles, Chapters, or Sections in the United States Code)?
	\item How do we define the \emph{length} of a single element?
	\item Given the length of an element (and potentially that of other elements), how do we decide whether it is \emph{long}?
\end{enumerate}

The answer to the first question depends on the context in which the method is deployed,
e.g., the highest level at which a lawmaker using the method can amend legal rules.
Therefore, we propose a generic long element detection method that works for all element types,
provided that for each element of a given type, we have access to the text nested inside it
(i.e., its own text and the text contained in its hierarchical descendants).

To answer the second question,
we define the length of an element to be the number of tokens contained in its text.
Notable alternatives to this definition of length are the number of non-whitespace characters,
the overall number of characters,
or (unless we are dealing with leaves in the element hierarchy) the number of children or descendants of an element.

In answering the third question,
we propose two complementary views.
First, since legal texts are typically accessed at their sequence level (e.g., the section level in the United States Code),
for sequence-level element types, an \emph{absolute} length measure based on readability seems appropriate,
and we suggest to flag all elements of these types as \emph{long} if they exceed the length of a typical page (i.e., $500$ tokens).
Second, for elements of all types, we can additionally adopt a \emph{relative} length measure
based on the length distribution of all elements of the same type that share a common ancestor of a specified other type in the element hierarchy
(cf. Figure~\ref{fig:law-structure-types}---%
e.g., all Sections in the same Title of the United States Code).
Here, our approach is to flag an element as \emph{long}
if it is longer than a user-chosen number of tokens. 
This number can be chosen as an absolute threshold based on semantic considerations (e.g., by converting a number of pages to a number of tokens),
as a relative threshold based on the complementary cumulative distribution function (CCDF) of the relevant token length distribution, which reveals what percentage of elements (y-axis) are longer than a certain number of tokens (x-axis), 
or as a function (e.g., the maximum) of both.

\subsubsection{Ambiguous Syntax}
\label{sssec:methods:as}

Recall from Section~\ref{sssec:theory:as} that \emph{ambiguous syntax} is the use of logical operators,
control flow operators,
or punctuation in a way that leaves room for interpretation.
We classify this smell as static, grammar-related, and local (cf.~Figure~\ref{fig:law-smell-taxonomy}), with text-based detectability based on discovering known patterns (cf.~Section~\ref{subsec:methods:tools}).

Since the known patterns we are looking for are characterized by syntactic ambiguity,
syntactic parsers can be helpful only in the sense that parse tree ambiguity \emph{suggests} ambiguous syntax.
Instead, we therefore resort to Regex.
These Regex are designed to capture formulations that are likely to create syntactic ambiguity,
using operators, punctuation, and wild cards as building blocks.

Our Regex allow us to flag problematic sequences of tokens---%
e.g., sequences featuring \emph{and} and \emph{or} in close proximity,
or contextualized usages of mathematically redundant syntax like  \emph{and/or}---,
which can then be judged quickly by human experts.

\subsubsection{Large Reference Tree}
\label{sssec:methods:lrt}

Recall from Section~\ref{sssec:theory:lrt} that a \emph{large reference tree} is a reference tree whose edge set exceeds a given size $x$,
where a reference tree rooted at an element of law $r$ is a tuple
composed of the elements $V_r$ reachable from $r$ by following references (including $r$),
and a minimal set of reference edges $E_r$ such that each element of $V_r$ can be reached from $r$.
We classify this smell as static, structure-related, and non-local (cf.~Figure~\ref{fig:law-smell-taxonomy}),
with graph-based detectability based on hierarchy edges and reference edges (cf.~Section~\ref{subsec:methods:tools}).

To detect large reference trees, we use the representation of legal documents as directed multigraphs from \cite{coupette2021},
resolving all hierarchy edges and reference edges at the lowest possible level.
From the resulting directed multigraphs, we remove those nodes (and their incident edges) that contain more than a specified number of tokens. 
This effectively excludes reference trees that are unlikely to be constructed in practice 
(e.g., when a Section references a Chapter of the United States Code, lawtakers are unlikely to read through the entire Chapter to understand the referencing Section).
We replace each reference edge to a non-leaf node with edges to all of its descendants that directly wrap text,
and construct trees rooted at all elements of law which contain references.
This allows us to assess the minimum number of references a lawtaker must follow in order to fully resolve the content of the provision at its root 
(i.e., $n-1$ references if the tree has $n$ nodes),
provided they do not encounter any semantic stopping criteria along the way.

A \emph{long reference chain} is any path from the root of a reference tree to a leaf at depth $\ell$,
where $\ell > x$ for a user-specified maximum acceptable chain length $x$.
$x$~might lie around $3$ if chosen in accordance with common (but not uncontested) user interface design heuristics,
or around $6$ if we trust users of \emph{law} to be twice as patient as regular users.
The length of reference chains depends on the reference resolution strategy (e.g., if it is more like breadth-first search or depth-first search), 
regarding which we deliberately refrain from assumptions.
We can, however, learn about the \emph{best} possible \emph{worst} case by analyzing reference chains in \emph{shortest path trees}.

\subsubsection{Natural Language Obsession}
\label{sssec:methods:nlo}

Recall from Section~\ref{sssec:theory:nlo} that \emph{natural language obsession} is the representation of typed data as natural language text.
We classify this smell as static, data-related, and local (cf.~Figure~\ref{fig:law-smell-taxonomy}),
with text-based detectability based on discovering known or unknown patterns (cf.~Section~\ref{subsec:methods:tools}).

Observe that, due to how we define a phrase in Section~\ref{sssec:methods:dp},
natural language obsession leads to duplicated phrases.
However, this only holds, at least in part,
because we preprocess the input text for duplicated phrase detection to replace named entities with known patterns.
Therefore, the detection of natural language obsession and the detection of duplicated phrases can be viewed as mutually enhancing:
We can preprocess the input text for duplicated phrase detection using tools to detect natural language obsession via \emph{known} patterns,
and use the results from duplicated phrase detection to detect instances of natural language obsession involving \emph{unknown} patterns.
Here, the classification of patterns as known versus unknown can refer to not only to us as the method designers but also to the users of our methods,
who might want to integrate their background knowledge into the law smell detection process.

It is worth noting that, although NER tools seem to be made for the task of detecting instances of natural language obsession using known patterns,
they usually require preprocessing steps that do not work reliably on legal texts (e.g., sentence splitting or part-of-speech tagging),
or their decisions are intransparent (e.g., when they are based on pretrained neural networks).
Thus, Regex incorporating domain expertise, along with lists of known legal entities, are again our tool of choice for the discovery of known patterns.
By running duplicated phrase detection and natural language obsession detection
in alternation, 
and treating past false negatives and false positives as future tests,
we can then reveal unknown patterns, 
and iteratively improve or complete our Regex and named entity lists.

\section{Practice: What Law Smells Exist in the Wild?}
\label{sec:practice}

In this section, we demonstrate the utility of the methods laid out in Section~\ref{sec:methods} to detect the law smells described in Section~\ref{sec:theory}.
We focus on identifying the example smells from Section~\ref{subsec:theory:examples}
with the methods described in Section~\ref{subsec:methods:examples},
using the United States Code at the end of each year between $1998$ and $2019$ (inclusive) as an example.
Our input data is taken from \citet{coupette2021},
who transform the XML files provided by the Office of the Law Revision Counsel of the United States House of Representatives for the annual versions of each Title of the United States Code 
into files following the data model sketched in Section~\ref{subsec:methods:assumptions} \citep[for details on the data preprocessing, see the Supplementary Material of][]{coupette2021}.

\subsection{Duplicated Phrase}
\label{practice:dp}


As sketched in Section~\ref{sssec:methods:dp}, 
to detect duplicated phrases,
we leverage the \textsc{Dupex} algorithm introduced by \citet{coupette2021a}.
We run this algorithm on each Title in each year separately, 
using the same data preprocessing and parametrization as the authors 
(i.e., we replace selected named entities identified by Regex with placeholders and set the maximum number of failures until the algorithm stops to $10\,000$).
To facilitate the analysis of our results, we draw on the postprocessing suggested in the \textsc{Dupex}
paper (i.e., we hierarchically cluster the term vectors of long and frequent duplicated phrases by their cosine similarity with Ward linkage and use the resulting phrase ordering for our manual inspection).

Table~\ref{tab:dp} gives examples of duplicated phrases discovered in the United States Code in $2019$, 
focusing on duplicates occurring across multiple Titles (top) as well as duplicates specific to Title~6 (bottom). 
The general duplicated phrases we identify often contain language related to scoping and definitions (e.g., ``the term \{term\} means'') or boilerplate conferring varying degrees of authority or discretion (e.g., ``as may be necessary to carry out the''), 
whereas Title-specific duplicated phrases frequently contain named entities not replaced during preprocessing (e.g., ``{[}{[}director] of the] bureau of citizenship and immigration services'')
or topic-related boilerplate (e.g., ``natural disasters, acts of terrorism, (and$|$or) other man-made disasters'').

While some of the variation we observe is entirely grammatical (e.g., ``there (are$|$is) authorized to be appropriated'') 
or semantically inconsequential (e.g., ``for [the] purposes of this$\dots$''), 
other variation could be semantically meaningful (e.g., ``and$|$or'' variation, 
as seen in ``cyber threat indicators (and$|$or) defensive measures''), 
raising the question whether all nuances in meaning are indeed intended.
Moreover, many of the legibility and maintainability challenges created by duplicated phrases could be easily mitigated through straightforward refactorings.
For example, the last duplicated phrase from Table~\ref{tab:dp} could be eliminated in three steps:
\begin{enumerate}
	\item Introduce \emph{ISE} as an abbreviation for \emph{Information Sharing Environment}.
	\item Define the term \emph{ISE information} as \emph{information within the scope of the information sharing environment, including homeland security information, terrorism information, and weapons of mass destruction information}.
	\item Use the newly introduced term to replace all occurrences of its definition.
\end{enumerate}

\begin{table}[t]
	\centering
	\caption{%
		Examples of duplicated phrases identified in multiple Titles of the United States Code in $2019$ (top) and in the $2019$ version of \emph{Title 6---Domestic Security} only (bottom), 
		with options and alternatives written in Regex syntax.
		For the general phrases,
		we additionally report the absolute and relative maximum occurrence frequency, 
		$\text{abs}_{\max}$ and $\text{rel}_{\max}$, 
		of each listed phrase across all Titles, 
		along with the Title $T$ in which it occurs.
		For the specific phrases, 
		we instead report the minimum and maximum occurrence frequency of all options or alternatives, 
		$c_{\min}$ and $c_{\max}$.
	}
	\label{tab:dp}
\begin{tabular}{p{0.715\linewidth}r}\toprule
	\bfseries General& $\text{abs}_{\max}~(T)$, $\text{rel}_{\max}~(T)$\\\midrule
	for [the] purposes of this (chapter$|$subchapter$|$section$|$subsection$|$paragraph)	&$4\,280~(26)$, $2.90~(26)$\\
	{[}except] as (defined$|$provided) (by$|$in) \{reference\}&$706~(42)$, $0.55~(11)$\\
	there (are$|$is) authorized to be appropriated&$1\,009~(42)$, $0.55~(34)$\\
	as may be necessary to carry out the&$133~(16)$, $0.10~\phantom{0}(2)$\\
	the term \{term\} means&$4\,438~(42)$, $2.38~(26)$\\
	\midrule
	\bfseries Specific&$c_{\min}$, $c_{\max}$\\
	\midrule
	{[}{[}director] of the] bureau of citizenship and immigration services&$18$, $66$\\
	natural disasters , acts of terrorism , (and$|$or) other man - made disasters&$11$, $52$\\
	\setlength\hangindent{0.5em}the committee on homeland security (and governmental affairs of the senate$|$of the house of representatives)&$30$, $31$\\
	cyber threat indicators (and$|$or) defensive measures&$20$, $26$\\
	\setlength\hangindent{0.5em}information within the scope of the information sharing environment , including homeland security information , terrorism information , and weapons of mass destruction information&$26$\\
	\bottomrule
\end{tabular}
\end{table}

\begin{figure}[t]
	\centering
	\includegraphics[width=\linewidth]{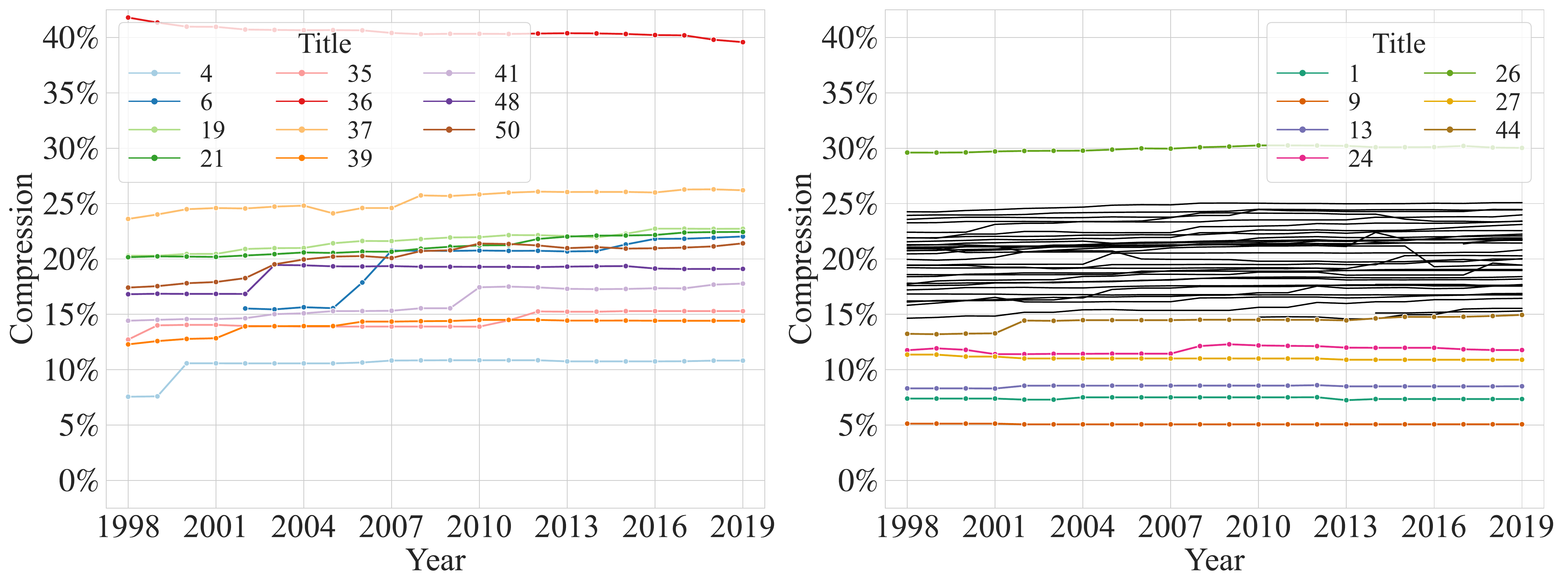}
	\caption{%
		Evolution of the compression achieved by the \textsc{Dupex} algorithm for different Titles of the United States Code ($1998$--$2019$), 
		in percent of the original encoded bit length.
		The left panel shows only Titles with a range of compression values of at least $2\%$.
		The right panel shows the remaining Titles, 
		colorizing those that constantly have a comparatively high compression (more than $25\%$) or a comparatively low compression (less than $15\%$). 
	}
	\label{fig:dp}
\end{figure}

\textsc{Dupex} extracts duplicated phrases by \emph{exploiting} redundancies to compress the input text.
Hence, we can use the compression it achieves to \emph{measure} the redundancy in this text. 
This allows us to track the development of redundancy---i.e., an upper bound on the \emph{deduplication potential}---%
in the Titles of the United States Code over time.
As shown in the right panel of Figure~\ref{fig:dp},
the deduplication potential of most Titles ranges between $15\%$ and $25\%$, 
with \emph{Title 9---Arbitration} (low deduplication potential) and \emph{Title 26---Internal Revenue Code} (high deduplication potential) as notable exceptions.
However, the left panel of Figure~\ref{fig:dp} highlights that the highest deduplication potential lies in \emph{Title 36---Patriotic and National Observances}, 
although this is one of the few Titles whose deduplication potential has \emph{decreased} over time.
In the left panel, we also see that the Title with the by far largest \emph{range} of deduplication potentials is \emph{Title 6---Domestic Security}, 
whose deduplication potential rises dramatically from $2005$ to $2007$. 
This increase follows the addition of Chapters~2 and 3 (in $2006$) and of Chapters~4 and 5 (in $2007$), 
which demonstrates how duplicated phrase detection can help pinpoint potentially problematic legislative acts.

\subsection{Long Element}
\label{practice:le}

\begin{figure}[t]
	\centering
	\includegraphics[width=\linewidth]{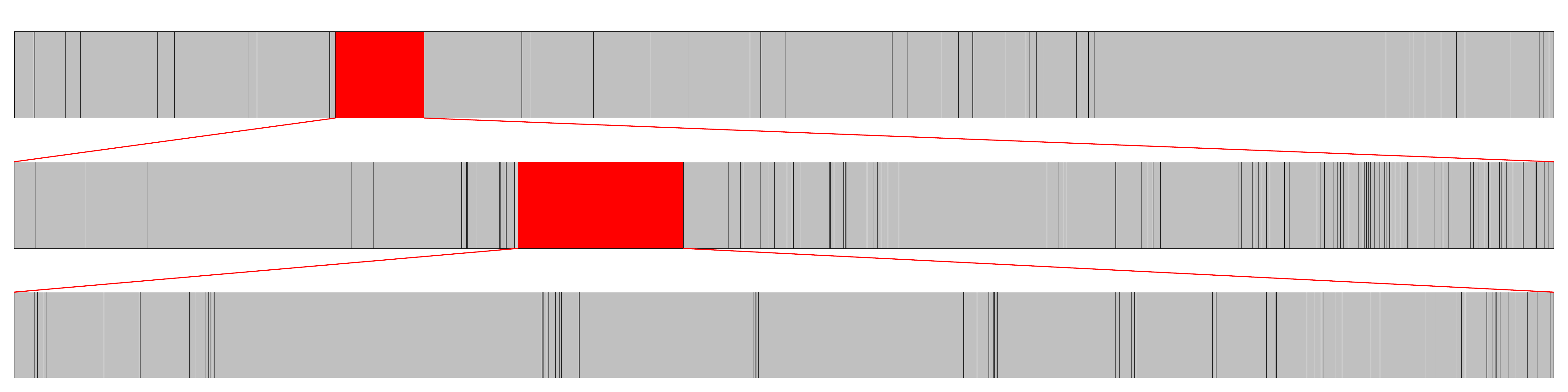}
	\caption{Zoomable icicle plot enabling visual element length assessment at
		different hierarchical levels of the $2019$ United States Code: Titles (top), Chapters (middle), and Sections (bottom). 
		Here, we zoom in on Chapters from Title 15 (Commerce and Trade) 
		and Sections from \uscchap{15}{14A} (Aid to Small Business).}
	\label{fig:le}
\end{figure}

As discussed in Section~\ref{sssec:methods:le}, 
many judgment calls need to be made to define what constitutes long elements in legal documents. 
Eager to avoid making these calls blindly, 
we propose zoomable icicle plots to interactively explore the element length distribution of hierarchically structured legal documents.
Figure~\ref{fig:le} depicts a static version of such a graphic, 
showing the Titles of the United States Code in $2019$ at the top, 
then zooming in on \emph{Title~15---Commerce and Trade} and its Chapters, 
and finally focusing on \emph{Chapter 14A---Aid to Small Business} and its Sections. 
As all horizontal space on layer $x$ (the \emph{higher} layer) is reclaimed on layer $x+1$ by the substructures of the selected element,
the element length distribution remains clearly discernible even on lower layers.
Thus, zoomable icicle plots allow us to intuitively identify long elements as candidates for refactoring, 
and to set length thresholds based on the lengths of these elements.

For a quantitative assessment, 
we focus on \emph{Sections} in the United States Code, whose length we measure in \emph{tokens}, 
and use a length \emph{ranking}, rather than a length \emph{threshold}, 
to explore our results.
Consequently, we list the ten longest Sections, along with their headings and lengths, in Table~\ref{tab:le},
and observe that the longest Section, \uscsec{42}{1395ww}, 
occupies roughly one hundred pages in standard print.
We further label the long Sections in  Table~\ref{tab:le} with their \emph{verbosity category} (i.e., the primary reason for their lengthiness, as apparent from manual inspection):
\emph{inline math} (i.e., verbal representations of computations and their parametrizations), \emph{listing} (i.e., verbal representations of tables, indices, or other enumerations), 
or \emph{substructure} (i.e., sections with elaborate substructure containing related rules).

Identifying the verbosity category of a long element can help us find applicable refactorings.
For example, 
\emph{inline math} could be outsourced to authoritative code that both \emph{documents} and \emph{implements} the mandated computation,
\emph{listings} could be outsourced to authoritative appendices and reformatted to better bring out their intrinsic structure, 
and elements with \emph{substructure} could be reshaped into a sequence of shorter elements.
Therefore, deriving a full, potentially more fine-grained taxonomy of verbosity categories constitutes an attractive avenue for further research.

\begin{table}[t]
	\centering
	\caption{The ten longest Sections of the United States Code in $2019$ (measured in tokens), along with the \emph{verbosity category} of their content.}
	\label{tab:le}
\begin{tabular}{r@{\hskip 2pt}c@{\hskip 2pt}l@{\hskip 5pt}p{0.52\linewidth}@{\hskip 5pt}r@{\hskip 5pt}l}
	\toprule
	\multicolumn{3}{c}{\bfseries Section}&\bfseries Heading&\bfseries Length&\bfseries Category\\
	\cmidrule{1-6}
	&42 U.S.C. &§~1395ww &Payments to hospitals for inpatient
	hospital services&50.3~K
	&Inline Math\\
	&16 U.S.C. &§~1274 &Component rivers and adjacent lands&35.4~K
	&Listing\\
	&42 U.S.C. &§~1396a &State plans for medical
	assistance&34.6~K
	&Substructure\\
	&42 U.S.C. &§~1395w-4 &Payment for physicians’ services&34.5~K
	&Inline Math\\
	&42 U.S.C. &§~1395l&Payment of benefits&29.0~K
	&Inline Math\\
	&15 U.S.C. &§~636&Additional powers&28.1~K
	&Substructure\\
	&42 U.S.C. &§~1395x&Definitions&27.7~K
	&Listing\\
	&42 U.S.C. &§~1395m&Special payment rules for particular items and services&25.9~K
	&Inline Math\\
	&42 U.S.C. &§~1396b&Payment to States&24.9~K
	&Substructure\\
	&\phantom{4}8 U.S.C. &§~1182&Inadmissible aliens&23.6~K
	&Listing\\
	\bottomrule
\end{tabular}
\end{table}

\subsection{Ambiguous Syntax}
\label{practice:as}

\begin{figure}[t]
	\includegraphics[width=\linewidth]{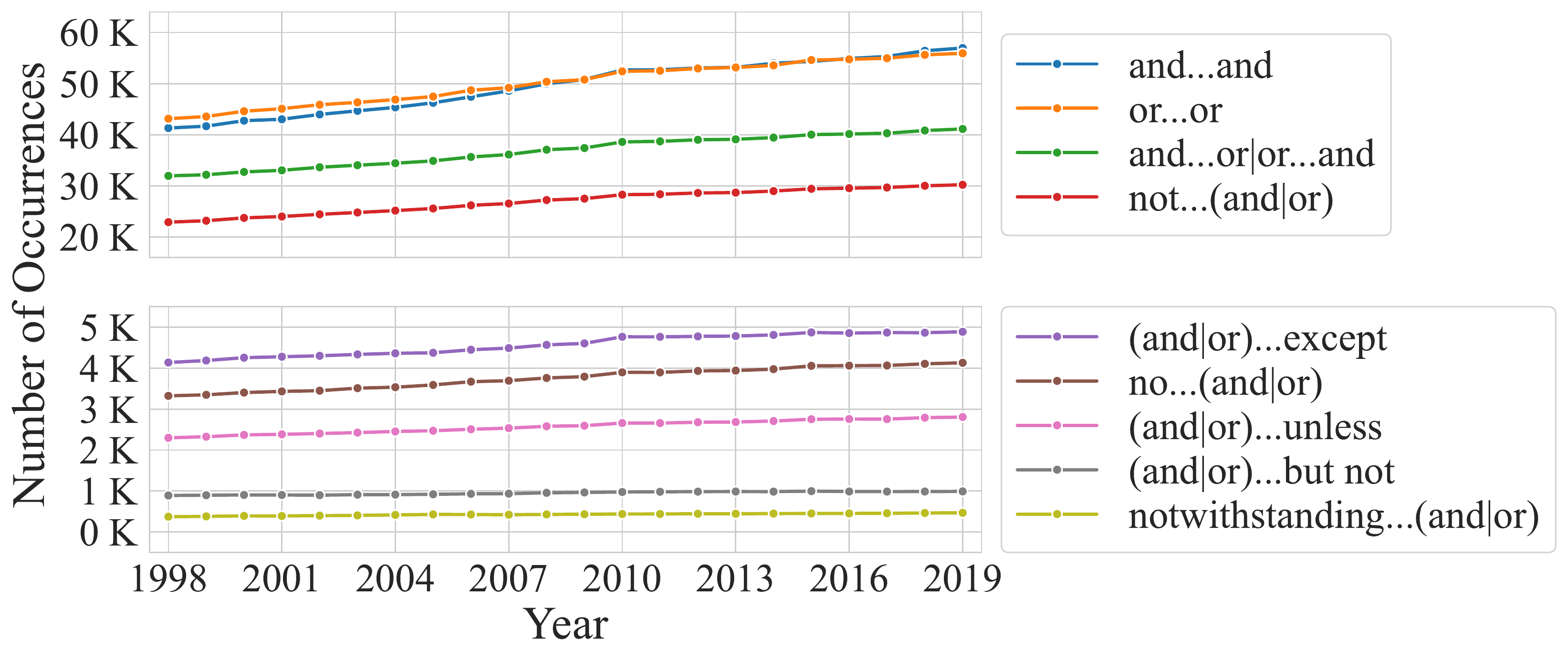}
	\caption{%
		Number of occurrences of ambiguous syntax candidates identified by problematic patterns in the United States Code  ($1998$--$2019$).}
	\label{fig:asabs}
\end{figure}

\begin{figure}[t]
	\includegraphics[width=\linewidth]{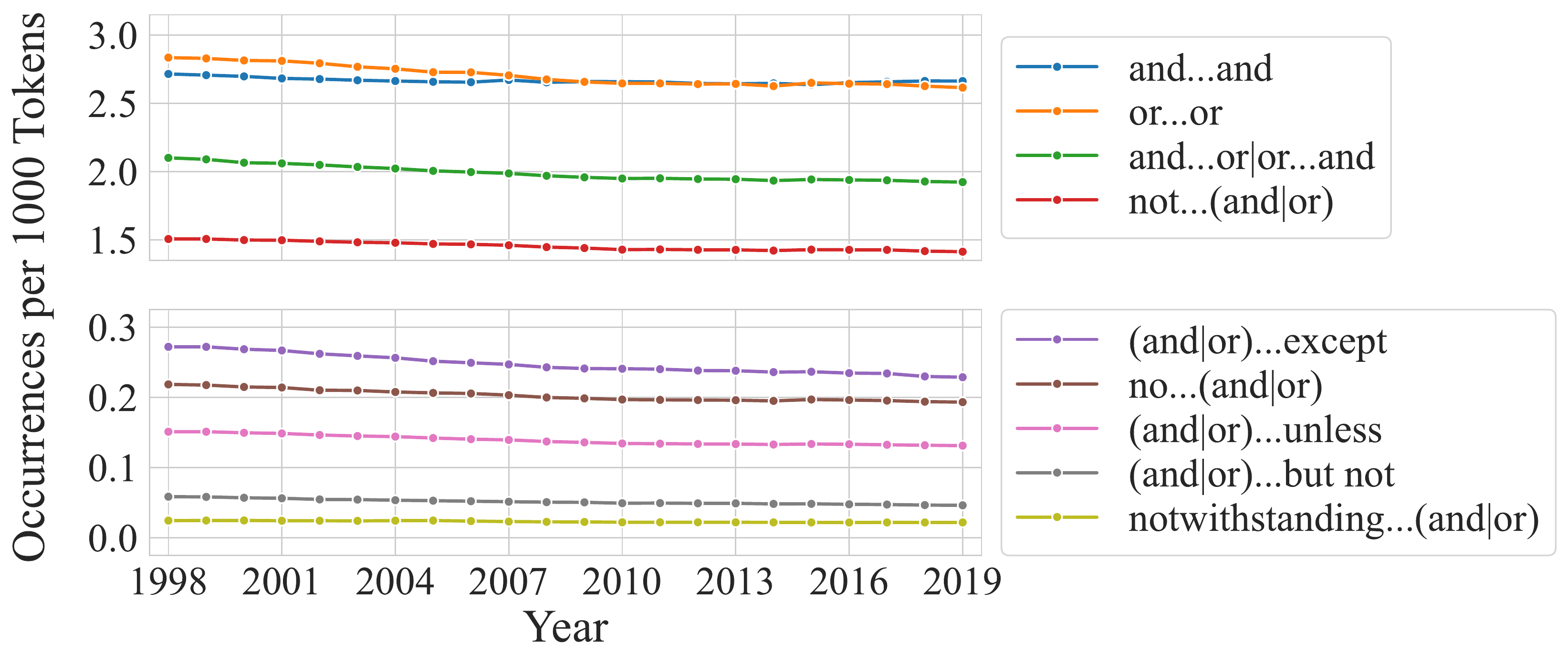}
	\caption{%
		Number of occurrences per $1\,000$ tokens of ambiguous syntax candidates identified by problematic patterns in the United States Code  ($1998$--$2019$).}
	\label{fig:asrel}
\end{figure}

As described in Section~\ref{sssec:methods:as},
we flag candidate instances of ambiguous syntax using Regex, 
leaving the final verdict regarding their ambiguity to a human reader (at least for the time being).
Here, we focus on three triples of \emph{problematic patterns} that revolve around the operators \emph{and} and \emph{or}.
Using ``$\dots$'' as a placeholder to denote \emph{at most fifty characters},
the first triple of patterns concerns multiple instances of \emph{and} or \emph{or} occurring together (\emph{and$\dots$and}, 
\emph{or$\dots$or}, 
\emph{and$\dots$or$|$or$\dots$and}),
the second triple involves the use of these operators with negation 
(\emph{no$\dots$(and$|$or)}, 
\emph{not$\dots$(and$|$or)}, 
\emph{notwithstanding$\dots$(and$|$or)}),
and the third triple combines them with oppositional conjunctions
(\emph{(and$|$or)$\dots$but not}, 
\emph{(and$|$or)$\dots$except}, 
\emph{(and$|$or)$\dots$unless}).
When tracking the absolute and relative prevalence of these patterns in the United States Code over time, 
as depicted in Figure~\ref{fig:asabs} and Figure~\ref{fig:asrel}, 
we observe a substantial and growing number of ambiguous syntax candidates, 
whose frequency remains relatively constant over time---%
despite personnel changes and technological advances.

We suspect that the set of ambiguous syntax candidates we generate using the Regex representing our problematic patterns is over-inclusive:
For example, when we find \emph{and} and \emph{or} in close proximity, 
the operator binding is often clarified by the context.
To estimate the degree of over-inclusiveness in our candidate set, 
we randomly sample $100$ candidates flagged by the \emph{and$\dots$or} pattern. 
In $38$ sampled cases, the operator binding is unclear from a grammatical point of view, 
i.e., deciphering the meaning of the affected phrase requires nontrivial background knowledge.
This demonstrates that instances of ambiguous syntax \emph{can} be identified in a computer-aided process, 
although more work needs to be done to exclude false positives (e.g., by refining our Regex or filtering flagged candidates).
 
To give a concrete example, 
one of our randomly sampled candidates is the second sentence of  \emph{\uscsec{12}{5538(a)(1)}---Mortgage loans; rulemaking procedures; enforcement} (emphasis added):
\begin{quote}
	Such rulemaking shall relate to unfair \textbf{or} deceptive acts \textbf{or} practices regarding mortgage loans, which may include unfair \textbf{or} deceptive acts \textbf{or} practices involving loan modification \textbf{and} foreclosure rescue services.
\end{quote}
The desired operator binding is \emph{probably}:
\begin{quote}
	Such rulemaking shall relate to (unfair \textbf{or} deceptive) (acts \textbf{or} practices) regarding mortgage loans, which may include (unfair \textbf{or} deceptive) (acts \textbf{or} practices) involving (loan modification \textbf{and} foreclosure rescue services).
\end{quote}
Here, we performed the simplest possible refactoring for instances of ambiguous syntax: adding brackets.
Without brackets, alternative readings of our sample sentence are possible---%
and the sentence is also much harder to read.

\subsection{Large Reference Tree}
\label{practice:lrt}

\begin{figure}[t]
	\centering
	\includegraphics[width=\linewidth]{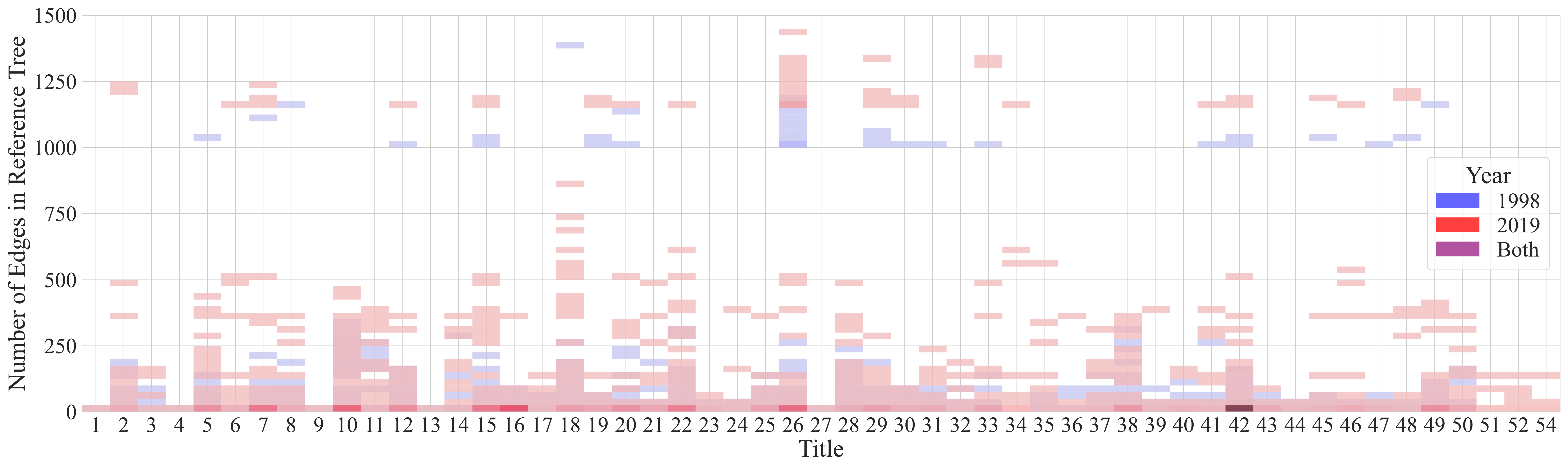}
	\caption{%
		Distribution of reference tree sizes across Sections of the United States Code in $1998$ and in $2019$.
		The two-dimensional histograms depict the density of Sections across Titles (x-axis) and reference tree sizes (y-axis).
	}\label{fig:lrt}
\end{figure}

As delineated in Section~\ref{sssec:methods:lrt}, 
we detect large reference trees in legal documents on the basis of their directed multigraph representation as introduced by \citet{coupette2021}, 
excluding overly broad references by removing all nodes that contain more than a specified number of tokens. 
For our exploratory purposes, we set this number to $1\,000$ tokens, 
i.e., roughly the amount of content that can be consumed on a large monitor without scrolling,
but other choices may be reasonable depending on the context.
We compute reference trees for all Sections of the United States Code from $1998$ to $2019$.
In Figure~\ref{fig:lrt}, we depict the distribution of reference tree sizes 
(i.e., their number of edges) across the different Titles in $1998$ and $2019$ as two-dimensional density histograms, 
where we include edges leading to nodes that have already been visited 
(i.e., edges closing cycles in the underlying undirected graph),
following the intuition that lawtakers might not keep track of this information.
These histograms not only show that Sections with relatively large reference trees exist in practice;
when overlaid as in Figure~\ref{fig:lrt}, 
they also reveal a shift towards larger reference trees in most Titles from $1998$ to $2019$, 
especially in the tail of the size distribution.
This extends prior findings that the law is growing increasingly interconnected \citep{katz2020,coupette2021}:
Even if we exclude overly broad references, 
thus modeling that they are unlikely to be resolved completely, 
understanding a Section of the United States Code in its context appears to become more laborious over time.

\begin{figure}[t]
	\centering
	\includegraphics[width=\linewidth]{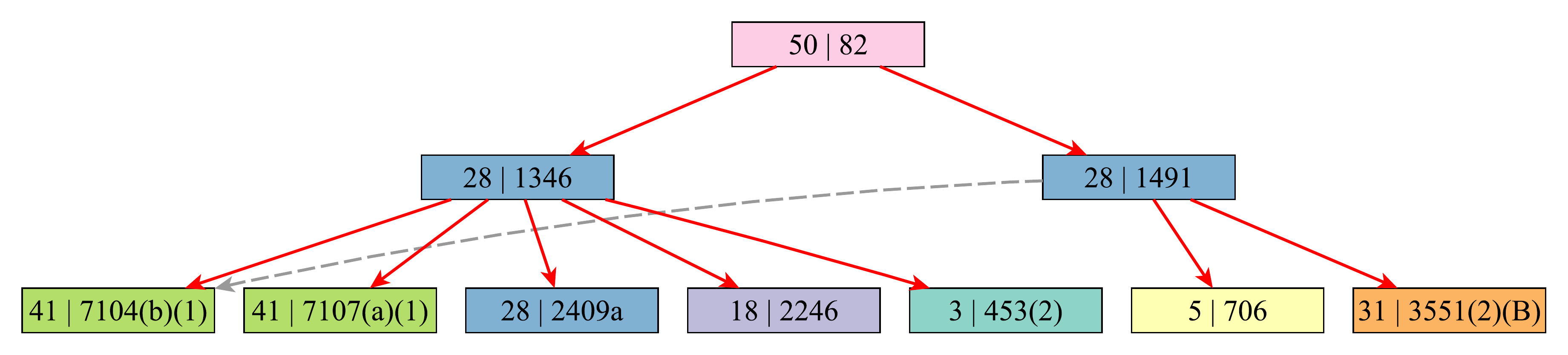}
	\caption{%
		The first two layers of the reference tree of \uscsec{50}{82} in $2019$.
		Nodes are labeled with ``\{Title\} $|$ \{Section[Subsection]\}'', 
		and node colors correspond to Titles.	
	}
	\label{fig:lrt-example}
\end{figure}

The reference tree of a Section $r$ in the United States Code can be measured in many dimensions:
Apart from its \emph{size} (i.e., its number of edges $|E_r|$),
we can assess, e.g., 
its path lengths---including, most notably, its \emph{depth} (i.e., the maximum length of a path from $r$ to any other node in its reference tree)---, 
its \emph{weight} (i.e., its total volume of text), 
and its \emph{diversity} (e.g., number of distinct Sections or Titles from which it contains nodes).
As \emph{large} reference trees are hard to depict statically, and \emph{deep} or \emph{heavy} reference trees are often also large,
for concreteness, 
we show a particularly \emph{diverse} reference tree in Figure~\ref{fig:lrt-example}.
Here, we resolve the first two layers of (United States Code-internal) references 
from \emph{\uscsec{50}{82}---Procurement of ships and material during war}, 
which takes us through seven different Titles in total (Title~50 included).

In the reference tree of \uscsec{50}{82}, 
all references from the root to the first layer are found in \emph{Subsection~(d)---Compensation for commandeered material}. 
This Subsection---a single sentence of more than $130$ tokens---not only exhibits a mixture of particularly unsavory law smells,
but it also addresses parties seeking compensation from the United States (i.e., the legislator), 
sending them to two Sections in Title~28 for details on what they can demand. 
To understand these details, potential claimants need to visit yet another five Titles on the second layer of the reference tree (in the worst case).
Our statistics show that this example is by no means exceptional,
which raises questions about the organization of the United States Code into Titles and the resulting user experience. 
As refactoring reference trees is nearly impossible without dedicated computational tools (due to the many interdependencies we need to account for),
the conceptualization and implementation of such tools is an important field for future work.

\subsection{Natural Language Obsession}
\label{practice:nlo}

As explained in Section~\ref{sssec:methods:nlo}, 
natural language obsession is closely related to duplicated phrases. 
First, we may use Regex as part of a preprocessing step to extract and replace instances of natural language obsession following \emph{known} patterns, 
which enables us to discover \emph{parametrized} duplicated phrases (as demonstrated in Section~\ref{practice:dp}).
Second, we can leverage the results of duplicated phrase detection to identify instances of natural language obsession following \emph{unknown} patterns.

Starting with the Regex-based approach, 
we extract text corresponding to data of the types \emph{money} (e.g., \emph{\$1,000}), 
\emph{percentage} (e.g., \emph{50 percent}), 
\emph{time period} (e.g., \emph{30 days}), 
and \emph{time point} (e.g., \emph{January~1}) 
for each Title of the United States Code from $1998$ to $2019$.
Above all, our statistics confirm that \emph{data as text} is ubiquitous in the United States Code: 
In $2019$, for example, we identify roughly 
19~K amounts of money, 
17~K percentages, 
35~K time periods, 
and 44~K time points.
However, these statistics come with three caveats:
First, our Regex are deliberately both simple and conservative, i.e., all numbers should be treated as lower bounds.
Second, the categories \emph{time period} and \emph{time point} group several time-related Regex, 
and the distinction between the classes is not always clearcut.
Third, when comparing annual counts across years, 
with or without normalization by annual Title size,
we can only \emph{observe} changes but not \emph{explain} them.

\begin{figure}[t]
	\centering
	\includegraphics[width=\linewidth]{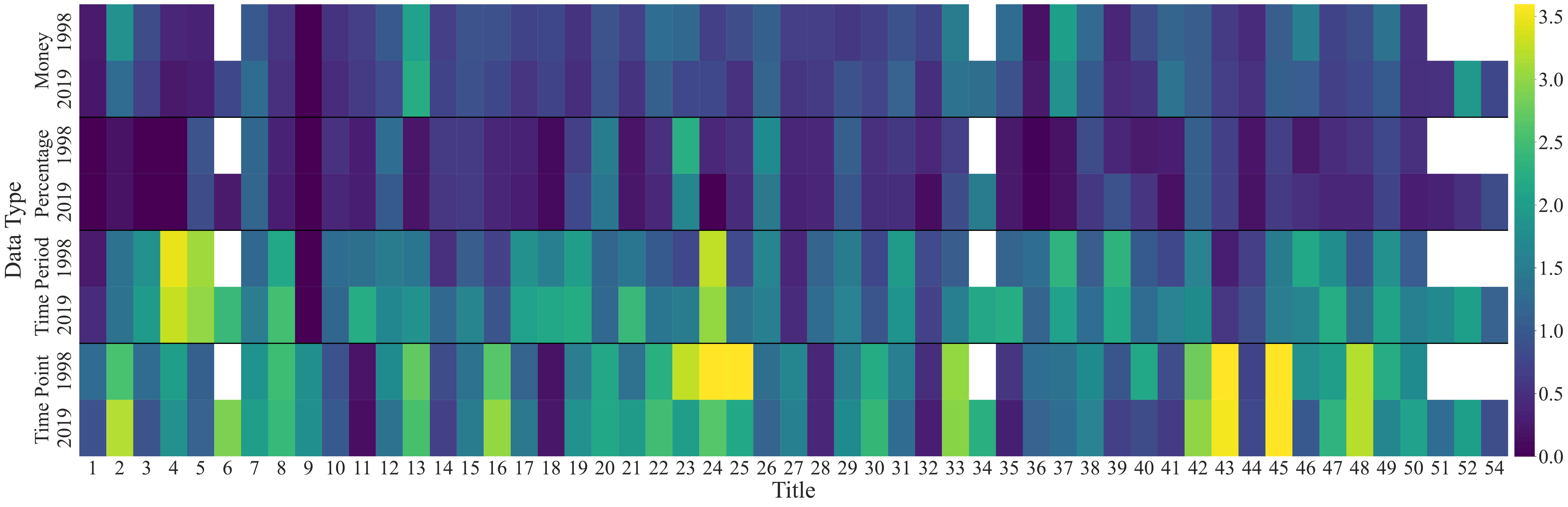}
	\caption{%
		Relative prevalence (instances per $1\,000$ tokens) of natural language obsession concerning data of the types \emph{money}, \emph{percentage}, \emph{time period}, and \emph{time point}, 
		for each Title in the United States Code in $1998$ and $2019$.
		The maximum of the color\-map is computed using the ninety-ninth percentile (rather than the most extreme value); 
		white fields signal that the corresponding Title has no tokens in the given year (Titles 6 and 34 had been repealed and not repurposed in $1998$, 
		and Titles 51, 52, and 53 did not even exist).
	}
	\label{fig:nlo-data}
\end{figure}

Figure~\ref{fig:nlo-data} depicts how many instances of the data types described above we detect per $1\,000$ tokens in each Title of the United States Code in $1998$ and $2019$.
With due caution, we make the following observations.
First, some Titles have distinct \emph{data profiles}, 
e.g., \emph{Title 24---Hospitals and Asylums} is unique in its comparatively high reliance on time periods \emph{and} time points.
Similarly, individual data types are particularly prevalent in a small set of Titles, 
e.g., in $2019$, \emph{Title 13---Census}, \emph{Title 37---Pay and Allowances of the Uniformed Services}, and \emph{Title 52---Voting and Elections}
have the highest frequency of money mentions.
Finally, the relative prevalence of certain data types increases between $1998$ and $2019$ for some Titles, 
while it decreases for others.
For example, \emph{Title 2---The Congress} states relatively more on time points in $2019$ than in $1998$, whereas \emph{Title 25---Indians} does the opposite.

\begin{figure}[t]
	\centering
	\includegraphics[width=\linewidth]{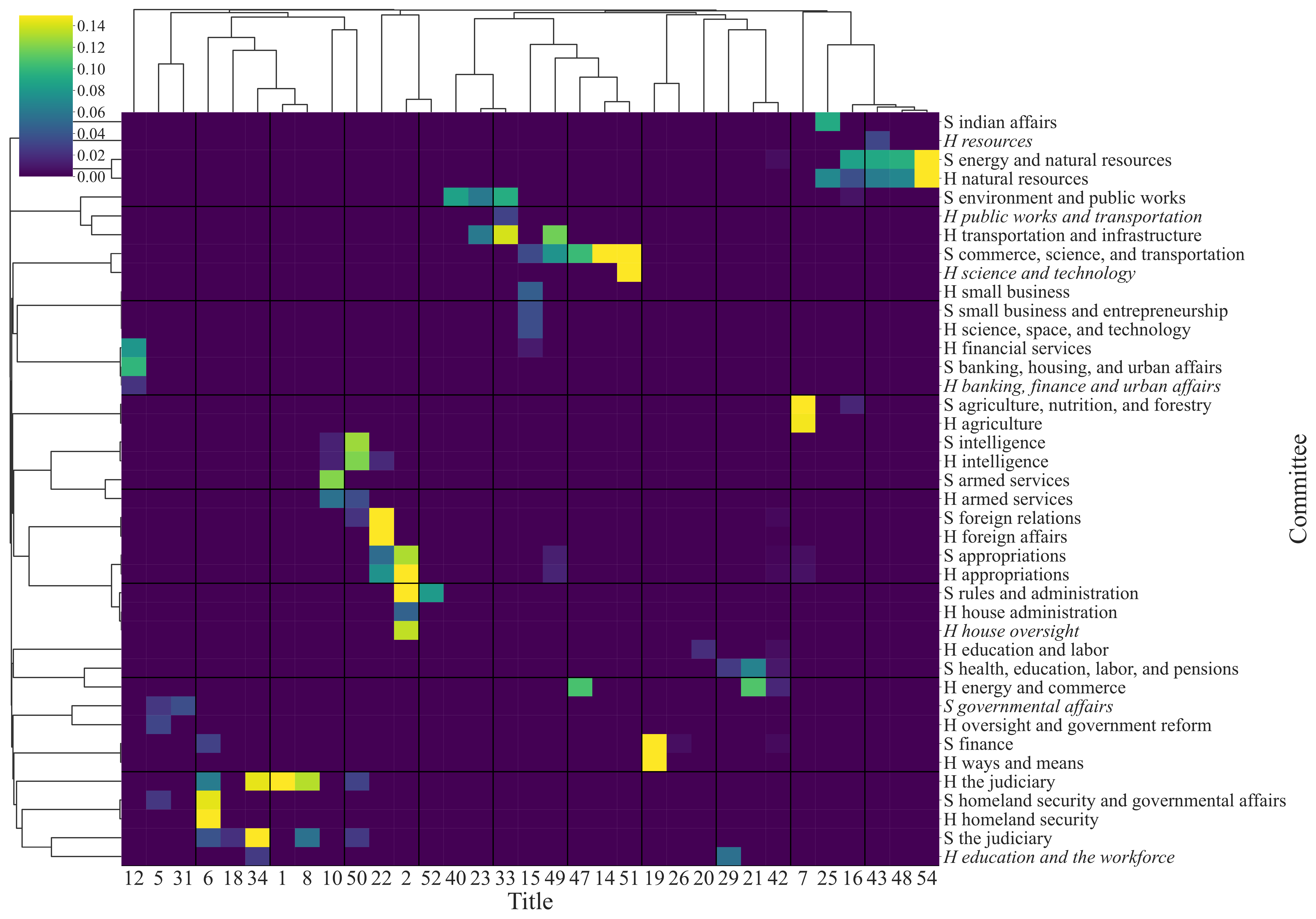}
	\caption{%
		Committee profiles for the Titles of the United States Code in $2019$.
		We depict the number of times per $1\,000$ tokens 
		that a Committee is mentioned with a name of the shape ``Committee on \{topic\} of the \{parent body\}'' in the duplicated phrases we extract for a Title, 
		where the maximum of the colormap is again computed using the ninety-ninth percentile. 
		The data are hierarchically clustered using correlation as a distance metric and average linkage.
		We omit Committees not contained with a name of the above-stated shape in a duplicated phrase for any Title, 
		and Titles in which we do not identify any Committee in a duplicated phrase.
		Committee names are abbreviated by the first letter of their parent body (\emph{Senate} or \emph{House of Representatives}) and their topic,
		and Committees that do not exist anymore are typeset in \emph{Italics}.
	}
	\label{fig:nlo-committees}
\end{figure}

Further illustrating the interplay between natural language obsession and duplicated phrases, 
we now turn to the instances of the former revealed by detecting the latter. 
We focus on two related types of named entities: 
Committees of the United States Senate and Committees of the United States House of Representatives. 
Figure~\ref{fig:nlo-committees} shows \emph{Committee profiles}, 
i.e., frequency estimates of Committee mentions per $1\,000$ tokens, 
for those Titles in the $2019$ United States Code in which we find at least one Committee mention of the shape ``Committee on \{topic\} of the \{parent body\}'' as part of a duplicated phrase.
The statistics underlying Figure~\ref{fig:nlo-committees} are again lower bounds on the actual frequencies, 
as we only count Committee occurrences when they are specified in the format stated above and part of at least one duplicated phrase. 
Hierarchically clustering the data using correlation distance and average linkage, 
we observe that in most cases, Senate and House Committees with similar topics are grouped together. 
The clustering also brings out similarities between Titles. 
For example,
\emph{Title 16---Conservation}, 
\emph{Title 43---Public Lands}, 
\emph{Title 48---Territories and Insular Possessions}, 
and \emph{Title 54---National Park Service and Related Programs} 
form a group (on the right) that has very low ``Committee correlation distance'', 
chiefly due to their frequent mentions of the \emph{Committee on Energy and Natural Resources of the Senate} and the \emph{Committee on Natural Resources of the House of Representatives}.

We find that almost all Standing Committees of the 116th Congress (active in $2019$) are part of at least one duplicated phrase,
the most notable exceptions being the \emph{Budget Committees} of the Senate and the House of Representatives 
(which are mentioned only in \emph{Title 2---The Congress}, often together, and mostly in abbreviated forms). 
Furthermore, not all of the ``Committees'' listed in Figure~\ref{fig:nlo-committees} existed in the 116th Congress.
For example, the \emph{Committee on Resources of the House of Representatives}, 
which we find in the duplicated phrases of Title~43, 
has been the \emph{Committee on Natural Resources} since $2007$ 
(returning to the name it had been carrying from $1993$ to $1995$).
\emph{Committee fossils} like this are widespread, 
especially among the House Committees,
and we typeset them in \emph{Italics} in Figure~\ref{fig:nlo-committees}.
They allow us to conclude with a textbook example highlighting the perils of natural language obsession: 
In the $2019$ version of Title~5, 
both \emph{Committee on Governmental Affairs of the Senate} and \emph{Committee on Homeland Security and Governmental Affairs of the Senate} are among our duplicated phrases, 
each with ten mentions.\!\footnote{%
	Note that these duplicated phrase counts do not necessarily correspond to the raw counts of the relevant phrase in the document of interest, 
	as the \textsc{Dupex} algorithm computes a \emph{cover} of the input token sequence, i.e., enforces that patterns are \emph{non-overlapping}.
} 
All of these mentions reference the same entity, 
i.e., what was known under the name \emph{Committee on Government Operations} from $1952$ to $1977$, 
reorganized as the \emph{Committee on Governmental Affairs} in $1977$,
and renamed as the \emph{Committee on Homeland Security and Governmental Affairs} in $2005$.
Upon closer inspection, we find additional mentions of the \emph{Committee on Governmental Affairs of the Senate}, 
in several name variants,
also in other Titles of the $2019$ United States Code (most notably in Title~31, cf. Figure~\ref{fig:nlo-committees}). 
As these are references to an entity that was renamed roughly fifteen years ago, 
this state is dissatisfying. 
While one might argue that the problem could be (and commonly \emph{is}) addressed in practice by 
the \emph{Committee Name History} provided by Congress\footnote{\url{https://www.congress.gov/help/committee-name-history}.}
and proper interpretation, 
this solution is neither efficient nor user-friendly. 
Here, computational approaches provide a viable alternative: 
Once a Committee is represented as an \emph{entity}, rather than a mere string, 
changing its name in all Titles of the United States Code becomes as simple as ``refactor...rename''.

\section{Discussion}
\label{sec:discussion}

In Section~\ref{sec:practice}, 
we have shown that the law smells described in Section~\ref{sec:theory} can be detected by the methods developed in Section~\ref{sec:methods}---%
at least in the United States Code from $1998$ to $2019$.
While we have focused on our five example law smells, i.e., 
\emph{duplicated phrases}, \emph{long elements}, \emph{large reference trees}, \emph{ambiguous syntax}, and \emph{natural language obsession}, 
it would be interesting to explore the prevalence of other law smells from our taxonomy in this particular corpus.
Moreover, our concepts and methods could immediately be applied to other corpora of statutes or regulations across countries, time periods, and regulatory levels,
such as the United States Code of Federal Regulations or the collection of European Union Regulations and Directives.
Leveraging law smells to monitor other collections of legal documents, 
e.g., judicial decisions or contracts, is less straightforward but also deserves further study.

Beyond its practical extensions, our work also leaves room for theoretical and methodological improvements. 
While our law smell taxonomy attempts to be comprehensive, it is most likely neither exhaustive nor ideally structured, 
and other smells and distinctions might prove useful in the future.
Likewise, our methodological toolkit is functional but by no means perfect.
For example, our Regex could easily integrate more background knowledge and cover more edge to reduce the number of false law smell negatives. 
Furthermore, we currently provide no interactive interfaces to investigate law smell detection results.

Finally, our work calls for extensions in three directions. 
First, we focus on law smell \emph{detection}, but our ultimate goal is law smell \emph{deodorization}.
To this end, we need both a better understanding of the landscape of \emph{legal refactorings} 
and better tools to perform those refactorings identified as applicable---%
in the best case, \emph{legal integrated development environments} combined with \emph{legal version control}.
Second, in many jurisdictions, there is currently no way for the legislator to change the language or structure of the law creating interpretive uncertainty concerning its content---%
i.e., refactoring in the classical software engineering sense is hardly possible and potentially undesirable,
and researchers could help legislators change this situation.
Third, we have deliberately skirted the questions \emph{how} and \emph{to which extent} law can be made computable,
which we consider an important research direction that could also build on some of our results.
Quests to answer this question might eventually lead us to adapt principles and ideas from object-oriented programming to the legal domain, 
which would yield many new law smells and corresponding refactorings. 

\section{Conclusion}
\label{sec:conclusion}

We have initiated the systematic study of \emph{law smells}, 
i.e., patterns in legal texts that might impair the comprehensibility and maintainability of the law 
and indicate the need for \emph{legal refactoring}.
Guided by five illustrative examples,
we have developed a comprehensive law smell taxonomy, 
introduced a methodological toolkit for law smell detection, 
and confirmed the utility of this toolkit in practice by applying it to $22$ years of codified statutory legislation in the United States Code ($1998$--$2019$).
Our work demonstrates how ideas from software engineering can be ported to the legal domain,
and it creates numerous opportunities for further research on \emph{defining}, \emph{detecting}, and \emph{deodorizing} law smells in \emph{all types of legal texts}.
Thus, we hope to have contributed a building block for the road towards truly computational legal drafting.




\section*{Data availability}
%
The data used in this study is archived under the following DOI: \tbd

\section*{Code availability}
%
The code used in this study is available on GitHub in the following repositories: \tbd

%
\section*{Conflict of interest}
The authors declare that they have no conflict of interest.


\bibliographystyle{spbasic}      
\bibliography{bibliography}   


\clearpage


\section*{\supp}

In this \supp, we detail the structured profiles of the example law smells introduced in Section~\ref{subsec:theory:examples}.

\subsection*{Duplicated Phrase}

\paragraph{Description}
A \emph{duplicated phrase} is a phrase above a specified \emph{length} that has more than a specified number of \emph{occurrences} in a legal text.
Here, a \emph{phrase} is a nonempty sequence of terms,
where a term is either a token (i.e., a sequence of non-whitespace characters roughly corresponding to a word)
or a placeholder for an argument that is itself a phrase.

\paragraph{Problem}
From a lawmaker's perspective, a duplicated phrase heightens the risk of inconsistencies 
when the phrase needs to be changed or is used to mean different things in different places.
From a lawtaker's perspective, a duplicated phrase reduces the information density of the text, making it harder to read,
and it increases the risk of confusion if the phrase is used with divergent meanings.

\paragraph{Detection}
Via scalable variants of $n$-gram search, possibly after some text preprocessing (e.g., to replace named entities).

\paragraph{Mitigation}
By introducing named variables and definitions.
Wherever a duplicated phrase occurs,
it can be replaced by its variable name and a reference to its definition
(which might be visible in the text or embedded in the markup).

\paragraph{Example}

The phrase ``information within the scope of the information sharing environment, including homeland security information, terrorism information, and weap\-ons of mass destruction information'' occurs $26$ times in the $2019$ version of Title $6$ of the United States Code.
From a parliamentary lawmaker's perspective, if $X$ is information that they consider ``within the scope of the information sharing environment'' (\emph{ISE information})
but a court decides that $X$ is \emph{not} ISE information, 
the lawmaker will need to add $X$ to the list of \emph{includes} after the first comma to get their will.
In its current structure, this requires $26$ changes to the text, whereas a single change will suffice if $X$ is maintained as a variable.
From a lawtaker's perspective, \emph{ISE information},
once defined by the duplicated phrase and then used consistently throughout,
would be much more readable and no less precise.
In the digital realm, the term could then be expandable to its full meaning, e.g., on click or on hover.

\subsection*{Long Element}

\paragraph{Description}
A \emph{long element} is an element containing legal text that is long as assessed by some absolute measure (e.g., number of tokens) or relative measure (e.g., quantile of a token length distribution).

\paragraph{Problem}
A long element may indicate a lack of structure in the legal text.
This increases the cognitive load for lawtakers reading the text,
and it complicates maintenance on the part of lawmakers.

\paragraph{Detection}
Via outlier detection using automatically computed length measures and (potentially nested) distributions for different types of elements (e.g., Titles, Chapters in a Title, or Sections in a Chapter of the United States Code).

\paragraph{Mitigation}
By moving (with the appropriate adjustments) some of their text to new elements or to shorter elements that already exist, i.e., by adding or altering structure.

\paragraph{Example}
In its $2019$ version, \uscsec{5}{552} contains more than $8\,000$ tokens,
i.e., more than $16$ normally typeset pages of text (or, more precisely: $22.5$ pages according to the page break markup in the official XML of the United States Code).
Its heading already indicates that the Section is a mixed bag: 
\emph{Public information; agency rules, opinions, orders, records, and proceedings}.%
\footnote{Computer scientists might be reminded of functions  with long names that include connectors,
	such as \emph{do\_x\_and\_y},
	which strongly indicate that these functions should be refactored.
}
This Section contains a large collection of rules related to public information,
organized in up to six levels of substructure
(small Latin alphabets, Arabic numbers, large Latin alphabets, small Roman numerals, large Roman numerals, small double Latin alphabets),
sometimes with more than $15$ substructure items on the same level.
From a lawmaker's perspective, a section of this length is hard to maintain in an orderly fashion.
For example, lawmakers seeking to add content related to \uscsec{5}{552} have incentives to simply append it to the section---%
there is not much order to ruin anyway, and integrating the new content into the section where it fits best would necessitate much more work.
This will make the section even longer, and even harder to maintain.
Since the substructure items have no headings,
lawtakers understandably unwilling to read \uscsec{5}{552} linearly are left to navigate it mostly by keyword search,
or---if they have the necessary background knowledge---by memory.
Breaking up \uscsec{5}{552} into several sections with separate headings would
allow them to restrict their search to potentially relevant text passages by making related, but relatively independent content directly accessible.

\subsection*{Ambiguous Syntax}

\paragraph{Description}
\emph{Ambiguous syntax} is the use of logical operators (e.g., \emph{and}, \emph{or}, and \emph{no(t)}),
control flow operators (e.g., \emph{if}, \emph{else}, or \emph{while}),
or punctuation (e.g., commas and semicolons) in a way that leaves room for interpretation.

\paragraph{Problem}
The intention of the legislator is communicated ambiguously,
which leads to legal uncertainty for lawtakers.
Eliminating that uncertainty further creates social costs (e.g., through lawsuits).

\paragraph{Detection}
Via pattern matching with regular expressions, followed by a manual assessment of the potentially problematic instances.

\paragraph{Mitigation}
By using logical operators with their precise mathematical meaning,
introducing \emph{xor} as a shorthand for the exclusive \emph{or} (to be clearly distinguished from the inclusive \emph{or}),
and adding brackets as syntax (e.g., round brackets to clarify operator binding, curly brackets to denote sets, rectangular brackets to denote lists).

\paragraph{Example}

Many sentences in the United States Code feature multiple instances of \emph{and} or \emph{or} as connectors.
For example, the second sentence of \uscsec{12}{5538~(a)~(1)} in $2019$ reads:
``Such rulemaking shall relate to unfair or deceptive acts or practices regarding mortgage loans, which may include unfair or deceptive acts or practices involving loan modification and foreclosure rescue services.''
Shall the referenced rulemaking relate to \emph{(((unfair or deceptive) acts) or (practices regarding mortgage loans))} or to \emph{((unfair or deceptive) (acts or practices)) regarding mortgage loans))}?
Can these loans include \emph{(((((unfair or deceptive) acts) or (practices involving loan modification)) and (foreclosure rescue services))} or \emph{(((unfair or deceptive) (acts or practices)) involving (loan modification and foreclosure rescue services))}?
Without clarifying syntax, from the text alone, we can assign higher or lower probabilities to the individual possibilities based on our estimates of intended sentence semantics,
but we cannot retrieve the accepted meaning without consulting external sources.

At the other end of the ambiguity spectrum,
the widespread uses of ``and/or'' (e.g., \uscsec{7}{451}: ``interstate and/or foreign commerce'')
and ``X, or Y[,] or both'' (e.g., \uscsec{26}{9012}: ``shall be fined not more than \$5,000, or imprisoned not more than one year or both'')
in the United States Code
are prime examples of redundant syntax leading to unnecessary verbosity (and distressed mathematicians).
Legally binding usage of \emph{or} for inclusive and \emph{xor} for exclusive options would allow lawmakers to replace ``and/or'' by ``or'' and do away with ``or both'', no ambiguities remaining.  

\subsection*{Large Reference Tree}

\paragraph{Description}
A \emph{reference tree} rooted at an element of law $r$ is a tuple $T_r = (V_r, E_r)$,
where $V_r$ is the set of elements of law reachable from $r$ by following references (including $r$),
and $E_r$ is a minimal set of edges (references)
such that each element of $V_r$ can be reached from $r$.
A reference tree is \emph{large} if its edge set exceeds a given \emph{size}.

\paragraph{Problem}
From a lawmaker's perspective, large reference trees may lead to unforeseen normative side effects, 
e.g., when a leaf element is changed.
From a lawtaker's perspective, large reference trees increase the cognitive load involved in navigating a legal text.

\paragraph{Detection}
By traversing adequately preprocessed directed multigraphs that represent legal document networks.

\paragraph{Mitigation}
By making references as specific as possible 
(i.e., not referencing a higher-level container if all referenced content is captured by a lower-level container),
or by restructuring the text and references contained in the tree elements 
(e.g., introducing lower-level containers that then can be referenced more precisely).

\paragraph{Example}
Many large reference trees can be found in tax law, e.g., in \emph{Title~26---Internal Revenue Code} of the United States Code.
For example, \uscsec{26}{62(a)(20)} references \uscsec{26}{62(e)}, which points to Sections spread across multiple Titles, 
and \uscsec{26}{751(c)(2)} references many Sections within Title~26.
More generally, provisions listing taxable and tax-exempted sources of income often point to definitions of these sources, 
which then point to further definitions and make further exceptions in their own text.
The resulting web of definitions, duties, and exemptions is hard to navigate for lay lawtakers, 
who need to rely on special-purpose applications (i.e., alternative user interfaces for parts of the United States Code) to determine their tax liability.

\subsection*{Natural Language Obsession}

\paragraph{Description}
\emph{Natural language obsession} is the representation of typed data as natural language text, 
often without a standardized format that could facilitate their algorithmic reconstruction.

\paragraph{Problem}
When represented using inconsistent natural language,
typed data is notoriously hard to parse and maintain.
Lawmakers forgo the benefits of type and consistency checking 
(e.g., if two interest rates stated in different parts of the text should always be equal),
lawtakers are left without semantic highlighting and expandable abbreviations,
and systematic analysis of typed data usage becomes impossible.

\paragraph{Detection}
Using Named Entity Recognition (NER) methods (including pattern matching with regular expressions),
potentially augmented or validated by scalable $n$-gram search techniques.

\paragraph{Mitigation}
By introducing a data layer that is separate from the text (representation) layer,
using strong types for named entities,
and associating type checking, highlighting, and data analysis rules with these types.

\paragraph{Example}
In the United States Code, punishments are often expressed using variants of the high-level pattern ``shall be fined not more than \{money\} or imprisoned not more than \{period\}'',
and temporal requirements of duties to act are frequently stated through variants of the high-level pattern ``not later than \{period\} after \{date\}''.
Reading, maintaining, and analyzing the rules following these patterns
(as a human or a computer) becomes much easier if the amounts of money, time periods, and dates are unambiguously identified as such,
follow a common format,
and are readily available for analysis in their natural units
(e.g., amounts of money as positive integers in national currency units or time periods as positive integers in units of days).
This can be achieved by storing typed data in a metadata layer supplementing the legal text, using their natural scales and units.

\end{document}